%% file: main_referee_report_response_1_arxiv.tex
\documentclass[preprint2,trackchanges,twocolumn]{aastex631}

\hypersetup{linkcolor=cyan, citecolor=cyan, filecolor=cyan, urlcolor=cyan}
\setcitestyle{semicolon, yysep={,}}

\shorttitle{Ly$\alpha$ and LyC escape in the Sunburst Arc}
\shortauthors{Owens et al.}

\graphicspath{{./}{figures/}}

\begin{document}

\title{Connecting Lyman-$\alpha$ and ionizing photon escape in the Sunburst Arc}

%% you desire. In v6.31 you can include a footnote in the title.

%% The new \altaffiliation can be used to indicate some secondary information
%% such as fellowships. This command produces a non-numeric footnote that is
%% set away from the numeric \affiliation footnotes.  NOTE that if an
%% \altaffiliation command is used it must come BEFORE the \affiliation call,
%% right after the \author command, in order to place the footnotes in
%% the proper location.
%%
%% While authors can be grouped inside the same \author and \affiliation
%% commands it is better to have a single author for each. This allows for
%% one to exploit all the new benefits and should make book-keeping easier.

%\correspondingauthor{August Muench}
%\email{greg.schwarz@aas.org, gus.muench@aas.org}

\author[0000-0002-2862-307X]{M. Riley Owens}
\correspondingauthor{M. Riley Owens}
\email{m.riley.owens@gmail.com}
\affiliation{Department of Physics, University of Cincinnati, Cincinnati, OH 45221, USA}

\author[0000-0001-6505-0293]{Keunho J. Kim}
\affiliation{Department of Physics, University of Cincinnati, Cincinnati, OH 45221, USA}
\affiliation{IPAC, California Institute of Technology, Pasadena, CA 91125, USA}

\author[0000-0003-1074-4807]{Matthew B. Bayliss}
\affiliation{Department of Physics, University of Cincinnati, Cincinnati, OH 45221, USA}

\author[0000-0002-9204-3256]{T. Emil Rivera-Thorsen}
\affiliation{The Oskar Klein Centre, Department of Astronomy, Stockholm University, AlbaNova, SE-10691 Stockholm, Sweden}

\author[0000-0002-7559-0864]{Keren Sharon}
\affiliation{Department of Astronomy, University of Michigan, 1085 S. University Ave, Ann Arbor, MI 48109, USA}

\author[0000-0002-7627-6551]{Jane R. Rigby}
\affiliation{Astrophysics Science Division, NASA Goddard Space Flight Center, 8800 Greenbelt Rd, Greenbelt, MD 20771, USA}

\author[0000-0001-7548-0473]{Alexander Navarre}
\affiliation{Department of Physics, University of Cincinnati, Cincinnati, OH 45221, USA}

\author[0000-0001-5097-6755]{Michael Florian}
\affiliation{Steward Observatory, University of Arizona, 933 North Cherry Ave., Tucson, AZ 85721, USA}

\author[0000-0003-1370-5010]{Michael D. Gladders}
\affiliation{Department of Astronomy \& Astrophysics, The University of Chicago, 5640 S. Ellis Avenue, Chicago, IL 60637, USA}
\affiliation{Kavli Institute for Cosmological Physics, University of Chicago, Chicago, IL 60637, USA}

\author[0000-0002-4519-7738]{Jessica G. Burns}
\affiliation{Department of Physics, University of Cincinnati, Cincinnati, OH 45221, USA}

\author[0000-0002-3475-7648]{Gourav Khullar}
\affiliation{Department of Physics \& Astronomy and PITT PACC, University of Pittsburgh, Pittsburgh, PA 15260, USA}

\author[0000-0002-0302-2577]{John Chisholm}
\affiliation{Department of Astronomy, The University of Texas at Austin, 2515 Speedway, Stop C1400, Austin, TX 78712, USA}

\author[0000-0003-3266-2001]{Guillaume Mahler}
\affiliation{Institute for Computational Cosmology, Durham University, South Road, Durham DH1 3LE, UK}
\affiliation{Centre for Extragalactic Astronomy, Durham University, South Road, Durham DH1 3LE, UK}

\author[0000-0003-2200-5606]{H\aa kon Dahle}
\affiliation{Institute of Theoretical Astrophysics, University of Oslo, P.O. Box 1029, Blindern, NO-0315 Oslo, Norway}

\author[0009-0000-6300-6184]{Christopher M. Malhas}
\affiliation{Department of Astronomy, The University of Texas at Austin, 2515 Speedway, Stop C1400, Austin, TX 78712, USA}

\author[0000-0003-1815-0114]{Brian Welch}
\affiliation{Astrophysics Science Division, NASA Goddard Space Flight Center, 8800 Greenbelt Rd, Greenbelt, MD 20771, USA}
\affiliation{Department of Astronomy, University of Maryland, College Park, MD 20742, USA}
\affiliation{Center for Research and Exploration in Space Science and Technology, NASA/GSFC, Greenbelt, MD 20771, USA}

\author[0000-0001-6251-4988]{Taylor A. Hutchison}
\affiliation{Astrophysics Science Division, NASA Goddard Space Flight Center, 8800 Greenbelt Rd, Greenbelt, MD 20771, USA}

\author[0009-0004-7337-7674]{Raven Gassis}
\affiliation{Department of Physics, University of Cincinnati, Cincinnati, OH 45221, USA}

\author[0000-0003-1343-197X]{Suhyeon Choe}
\affiliation{The Oskar Klein Centre, Department of Astronomy, Stockholm University, AlbaNova, SE-10691 Stockholm, Sweden}

\author[0009-0003-3123-4897]{Prasanna Adhikari}
\affiliation{Department of Physics, University of Cincinnati, Cincinnati, OH 45221, USA}

\collaboration{20}{(Sloan Giant Arcs Survey)}

\begin{abstract}

We investigate the Lyman-$\alpha$ (Ly$\alpha$) and Lyman continuum (LyC) properties of the Sunburst Arc, a $z=2.37$ gravitationally lensed galaxy with a multiply-imaged, compact region leaking LyC and a triple-peaked Ly$\alpha$ profile indicating direct Ly$\alpha$ escape. Non-LyC-leaking regions show a redshifted Ly$\alpha$ peak, a redshifted and central Ly$\alpha$ peak, or a triple-peaked Ly$\alpha$ profile. We measure the properties of the Ly$\alpha$ profile from different regions of the galaxy using $R\sim5000$ Magellan/MagE spectra. We compare the Ly$\alpha$ spectral properties to LyC and narrowband Ly$\alpha$ maps from Hubble Space Telescope (HST) imaging to explore the subgalactic Ly$\alpha-$LyC connection. We find strong correlations (Pearson correlation coefficient $r>0.6$) between the LyC escape fraction ($f_{\rm esc}^{\rm LyC}$) and Ly$\alpha$ (1) peak separation $v_{\rm{sep}}$, (2) ratio of the minimum flux density between the redshifted and blueshifted Ly$\alpha$ peaks to continuum flux density $f_{\rm{min}}/f_{\rm{cont}}$, and (3) equivalent width. We favor a complex \ion{H}{1} geometry to explain the Ly$\alpha$ profiles from non-LyC-leaking regions and suggest two \ion{H}{1} geometries that could diffuse and/or rescatter the central Ly$\alpha$ peak from the LyC-leaking region into our sightline across transverse distances of several hundred parsecs. Our results emphasize the complexity of Ly$\alpha$ radiative transfer and its sensitivity to the anisotropies of \ion{H}{1} gas on subgalactic scales. Large differences in the physical scales on which we observe spatially variable direct escape Ly$\alpha$, blueshifted Ly$\alpha$, and escaping LyC photons in the Sunburst Arc underscore the importance of resolving the physical scales that govern Ly$\alpha$ and LyC escape.

\end{abstract}



\section{Introduction} \label{sec:intro}

% Overivew of cosmological history
Following the formation of the first stars and galaxies, the universe experienced a phase transition during an era known as the Epoch of Reionization (EoR) ($z\sim 6-10$). Prior to this, protons and electrons sufficiently cooled after the hot Big Bang to recombine ($z\sim 1100$; the Epoch of Recombination) into neutral hydrogen (\ion{H}{1}). As the earliest generations of ionizing sources formed, they reionized this neutral gas.

% Problem of reionization and contribution of quasars
How the universe reionized remains an outstanding problem of modern astronomy. The two most likely sources of the ionizing radiation (called the Lyman continuum (LyC), with $\lambda<912$ {\AA}) are star-forming galaxies and active galactic nuclei (AGN). %The relative contributions of either group to reionization is unclear, as well as how those contributions evolved over time. On one hand, 
A drastic reduction in the AGN number density at $z>6$ \citep[][]{2012ApJ...755..169M, 2013A&A...551A..29P} might suggest AGN contribute $<10\%$ (but very likely $<30\%$) of the reionization `budget' \citep{2012ApJ...755..169M, 2017MNRAS.465.1915R}, but JWST's discovery of a faint AGN population during and shortly after reionization \citep{kocevski+2023,dayal+2024} could challenge this. Other authors have constructed AGN-dominated scenarios consistent with multiple observational constraints \citep{madau+2015}, so the role of AGN in reionization is still controversial. 

% Role of SFGs in reionization
If star-forming galaxies dominated the reionization of the universe, LyC photons from ionizing sources in star-forming galaxies must have escaped their host galaxy somehow. However, direct observations of the fraction of LyC which escapes galaxies ($f_{\text{esc}}^{\text{LyC}}$) become increasingly difficult at $z>2$, and nearly impossible at $z>4$ \citep{inoue+2014}. This is because the intergalactic medium (IGM) becomes increasingly neutral---and therefore increasingly opaque to LyC photons---as we look back beyond $z\sim2$. %. Since \ion{H}{1} rapidly attenuates LyC, direct observations of LyC during the EoR require a serendipitously ionized channel in the intergalactic medium (IGM) along the line-of-sight to the galaxy. 
%At larger redshifts, where less reionization has happened, this becomes more unlikely and effectively impossible.

Additionally, various cosmological hydrodynamical simulations disagree on how different sources contributed to reionization and on the dependence of $f_{\text{esc}}^{\text{LyC}}$ on important properties like galaxy mass and redshift \citep{2008ApJ...672..765G, ma+2015, 2016MNRAS.458L..94S, kannan+2022}. %The situation becomes further complicated because the typical $f_{\text{esc}}^{\text{LyC}}$ of galaxies in the local universe cannot explain the observed reionization timescale of hundreds of Myr.
The dilemma becomes further complicated because the observed timescale of reionization (hundreds of Myr; e.g., \citet{planck_collaboration2016}) is shorter than expected if the galaxies that caused reionization had similar $f_{\rm{esc}}^{\rm{LyC}}$ to local galaxies. This discrepancy suggests the LyC production and escape properties measured in local galaxies do not fully describe the processes that drove reionization. %adopting the typical $f_{\rm{esc}}^{\rm{LyC}}$ of local galaxies as representative of reionization galaxies would lead to reionization timescales much longer than the observed time of hundreds of Myr. %in sufficient quantities to explain the near-completely ionized (neutral fraction $\sim10^{-4}$ \citep{2006AJ....132..117F}) IGM.
If AGN do not dominate reionization, then to satisfy the reionization budget, $\gtrsim10-20\,\%$ of LyC must escape from source galaxies \citep[][]{2007MNRAS.382..325B, 2009ApJ...706.1136O, 2012MNRAS.423..862K, 2013ApJ...768...71R, 2015ApJ...802L..19R, 2015MNRAS.454L..76M, 2015ApJ...811..140B, 2016MNRAS.457.4051K, 2016arXiv160503970P}. But LyC surveys consistently find typical escape fractions $\lesssim10\%$ \citep[][]{2010ApJ...725.1011V, 2015ApJ...814L..10S, 2016A&A...585A..48G, 2017A&A...602A..18G, 2016A&A...587A.133G, 2016ApJ...819...81R, 2017ApJ...841L..27R, 2016ApJ...831...38V}. 
Although JWST has revealed that some galaxies produce ionizing photons much more efficiently than anticipated (e.g., \citet{atek+2023, endsley+2024}), which would permit reionization scenarios with lower typical escape fractions, it is unclear if these discrepant galaxies are sufficiently numerous. For a complete review of the current understanding of reionization and JWST's expected impact, see \citet{robertson+2022}.

One strategy to identify LyC production and escape methods is to observe post-reionization galaxies believed to be good analogues to reionization galaxies \citep{izotov+2011a, izotov+2016d, izotov+2017a, izotov+2017b, izotov+2018a, izotov+2019a, izotov+2019b, izotov+2021a, steidel+2018, fletcher+2019, flury+2022a, flury+2022b, pahl+2022, schaerer+2022b}. Targets at $z\lesssim4$ avoid the excessive IGM attenuation seen at higher redshift (though it is often still nontrivial), such that rest-LyC radiation is accessible with ultraviolet (UV) or optical filters on ground- and space-based instruments. The primary goal of observing later analogues to reionization galaxies is to link LyC escape to UV diagnostics of a galaxy's interstellar medium (ISM) and stellar populations, which together strongly regulate the production and escape of LyC photons. Connecting UV diagnostics and LyC is vital because the non-ionizing UV to rest-optical is the wavelength band JWST (and soon the generation of extremely large telescopes) directly accesses from reionization galaxies \citep{williams+2022, bunker+2023, lin+2023, mascia+2023}. The relations between UV diagnostics and $f_{\text{esc}}^{\text{LyC}}$ established in lower-redshift analogues can then constrain $f_{\text{esc}}^{\text{LyC}}$ for reionization galaxies and the primary escape methods of LyC photons.

The overarching strategy to find analogues to reionization galaxies is to preselect LyC-leaking candidates based on indirect tracers of LyC escape (e.g., the [\ion{O}{3}] 4959, 5007 {\AA} / [\ion{O}{2}] 3727, 3729 {\AA} ratio O32 or Ly$\alpha$ peak separation $v_{\text{sep}}$). In this fashion, many post-reionization LyC leakers have been identified \citep{bergvall+2006, leitet+2011, leitet+2013, borthakur+2014, mostardi+2015, deBarros+2016, izotov+2016a, izotov+2016b, izotov+2018c, izotov+2018d, izotov+2020, izotov+2021b, izotov+2022, leitherer+2016, shapley+2016, bian+2017, puschnig+2017, steidel+2018, vanzella+2018, fletcher+2019, ji+2020, saha+2020, jones+2021, malkan+2021b, marques-chaves+2021, marques-chaves+2022, saxena+2022, flury+2022a, flury+2022b}, though generally with insufficient spatial resolution to resolve the small scales (down to tens of parsecs and less) expected to dictate LyC escape. Notably, \citet{rivera-thorsen+2022} conducted a bottom-up search for LyC sources selected solely by their LyC emission in archival HST imaging that suggested the preselection criteria normally imposed in LyC emitter searches may miss nontrivial contributors to the ionizing background.

\begin{figure*}[ht!]
    \centering
    \includegraphics[width=\textwidth]{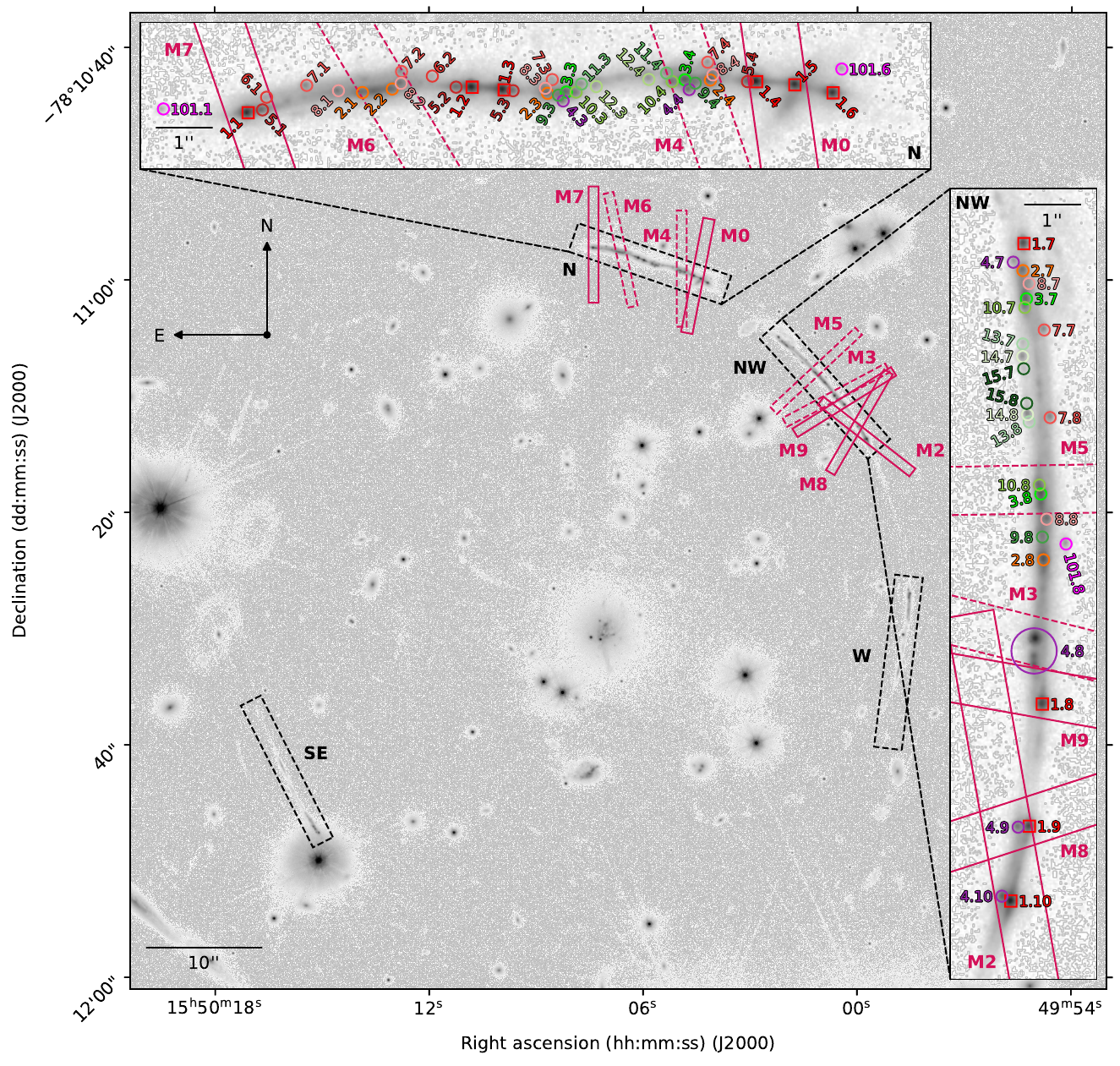}
    \caption{HST/WFC3 F606W imaging of the Sunburst Arc from GO-15377 (PI: M. Bayliss). Black dashed rectangles on the main image mark the main arc segments, labeled as in \citet{sharon+2022}. The inset panels enlarge the two largest arcs, which are the focus of spectroscopic MagE data presented in this work (Table \ref{tab:mage_log}). The MagE apertures appear as magenta rectangles, where solid apertures cover images of the LyC-leaking region, and dashed apertures do not. In the inset panels, circles (except for squares for the LyC-leaking region) mark images of different source plane regions of the Sunburst Arc, as labeled and colored in the lens model of \citet{sharon+2022}.}
    % \textit{Left}: the lensing cluster field containing the Sunburst Arc, with main arcs boxed in blue and labeled according to \citet{sharon+2022}. \textit{Right}: MagE apertures on the N and NW arcs (magenta) and images of the LyC-leaking region studied by \citet{rivera-thorsen+2019} (blue). MagE apertures in solid lines cover images of regions of the galaxy that leak LyC and dashed apertures do not cover any regions of the galaxy that leak LyC.}
    %\caption{HST/ACS F814W imaging of the Sunburst Arc from GO-15101 (PI: H. Dahle).  \textit{Left}: the lensing cluster field containing the Sunburst Arc, with main arcs boxed in blue and labeled according to \citet{sharon+2022}. \textit{Right}: MagE apertures on the N and NW arcs (magenta) and images of the LyC-leaking region studied by \citet{rivera-thorsen+2019} (blue). MagE apertures in solid lines cover images of regions of the galaxy that leak LyC and dashed apertures do not cover any regions of the galaxy that leak LyC.}
    \label{fig:arcs_mage}
\end{figure*}

% How the Sunburst Arc fits into this picture
This search for tracers of LyC escape, spurred by the simultaneous necessity to understand the process of reionization and inability to directly observe LyC from reionization, makes the galaxy PSZ1-ARC G311.6602-18.4624---nicknamed the \textit{Sunburst Arc} by \citet{rivera-thorsen+2017}---a compelling target to study LyC escape. The Sunburst Arc is a strongly lensed (magnifications between $\sim10-100\times$ \citep{rivera-thorsen+2019, 2021A&A...655A..81P, sharon+2022}) galaxy arc at redshift $z\approx2.37$. Discovered by \citealt{2016A&A...590L...4D}, the Sunburst Arc has garnered great interest because of (1) its high LyC escape fraction \citep{rivera-thorsen+2017, rivera-thorsen+2019, 2021arXiv210610280V}, (2) the unique mode of LyC escape implied by its Ly$\alpha$ emission, and (3) the advantages offered by the strong lensing in studying LyC escape. The foreground galaxy cluster PSZ1 G311.65-18.48 at $z\approx0.44$ lenses the Sunburst Arc into 12 full or partial images of the galaxy that show at least 54 individual features in total \citep{2021A&A...655A..81P, sharon+2022}. Some of these images could represent individual star clusters with radius as small as $\lesssim20$ pc \citep{2021arXiv210610280V}. A compact LyC-emitting region appears at least 12 times \citep{rivera-thorsen+2019}, and may be less than 10 pc across \citep{mestric+2023}.

Previous studies strongly disfavor the presence of an AGN in the Sunburst Arc \citep{rivera-thorsen+2019, mainali+2022}. Instead, past work favors young, massive stars produced in a recent burst of star formation as responsible for the observed LyC photons, mainly evidenced by the P-Cygni wind profiles, high-velocity and highly ionized galactic outflows, and compact regions with extremely blue UV slopes characteristic of hot, young stellar populations \citep{rivera-thorsen+2017, rivera-thorsen+2019, chisholm+2019, mainali+2022, mestric+2023, pascale+2023, kim+2023}. Together, the excellent spatial resolution in the source plane, multiple lines of sight into the galaxy, high signal-to-noise, and young stellar populations make the Sunburst Arc an ideal target to understand how LyC photons may have escaped from earlier star-forming galaxies during reionization.

\begin{deluxetable*}{ccccccccc}[t!]
    \tablecaption{MagE observation log and magnifications}
    \label{tab:mage_log}
    \tablehead{
        \colhead{Slit} &
        \colhead{Position} &
        \colhead{Width} &
        \colhead{Date} &
        \colhead{Exp. time} &
        \colhead{FWHM} &
        \colhead{R} &
        \colhead{$\mu$}
        \\
         &
        [(hh:mm:ss, dd:mm:ss)] &
        [$\arcsec$] &
         &
        [ks] &
        [$\arcsec$] &
         &
         &
    }
    \startdata
    M5 & (15:50:01.1649, -78:11:07.822) & 0.85 & 2018 Apr 22 & 13.5 & $0.97$ & 5500 $\pm$ 400 & $51 ^{+5}_{-10}$
    \\
    M4 & (15:50:04.9279, -78:10:59.032) & 0.85 & 2018 Apr 21 & 7.2 & $0.71$ & 5400 $\pm$ 300 & $14.6 ^{+0.8}_{-3}$
    \\
    M6 & (15:50:06.6389, -78:10:57.412) & 0.85 & 2018 Apr 22, 23 & 12.9 & $0.76$ & 5300 $\pm$ 300 & $147^{+5}_{-20}$
    \\
    M3 & (15:50:00.6009, -78:11:09.912) & 0.85 & 2018 Apr 21 & 12 & $0.70$ & 5500 $\pm$ 400 & $36 ^{+4}_{-5}$\tablenotemark{a}
    \\
    \hline
    M0 & (15:50:04.4759, -78:10:59.652) & 1 & 2017 May 24 & 13.5 & $1.34$ & 4700 $\pm$ 200 & $10 ^{+10}_{-7}$
    \\
    M2 & (15:49:59.7480, -78:11:13.482) & 0.85 & 2018 Apr 21, 23 & 8.1 & $0.77$ & 5300 $\pm$ 300 & $32 ^{+6}_{-3}$
    \\
    M7 & (15:50:07.3959, -78:10:56.962) & 0.85 & 2018 Aug 11, 12 & 14.4 & $0.73$ & 5200 $\pm$ 200 & $35 ^{+3}_{-6}$
    \\
    M8 & (15:49:59.9499, -78:11:12.242) & 0.85 & 2018 Aug 10, 11, 12 & 14.4 & $0.70$ & 5200 $\pm$ 300 & $29^{+6}_{-3}$
    \\
    M9 & (15:50:00.3719, -78:11:10.512) & 0.85 & 2018 Apr 23 & 13.5 &  $0.68$ & 5500 $\pm$ 400 & $31 ^{+4}_{-3}$
    \enddata
    \tablecomments{From left to right: slit label, position in (right ascension (hh:mm:ss), declination (dd:mm:ss)) (J2000), slit width in arcseconds, UT observation date, total exposure time in kiloseconds, individual exposure time-weighted average of the seeing conditions in arcseconds, median spectral resolution (R) as determined by the widths of night sky lines, and average magnification ($\mu$), as calculated using the lens model of \citet{sharon+2022}.}
    \tablenotetext{a}{Although slit M3 captures an extremely magnified region of the galaxy, the quoted magnification is puzzlingly non-extreme because the extreme magnification relies upon a postulated, unseen perturbing mass (likely detected in JWST/NIRCam imaging by \citet{choe+2024}), the effect of which is not included in the model used to calculate the magnifications listed here.}
\end{deluxetable*}

A promising tool to investigate LyC escape is a source's Lyman-$\alpha$ (Ly$\alpha$) emission (see, e.g., \citet{verhamme+2015}). Ly$\alpha$ photons primarily originate from recombining \ion{H}{2} ions and free electrons in \ion{H}{2} regions around massive OB stars producing copious LyC photons, implying that Ly$\alpha$ and LyC photons begin their journeys not far from each other. As a resonant emission line of \ion{H}{1}, Ly$\alpha$ photons interact strongly with the same gas that attenuates LyC from sources within a galaxy (stronger, in fact, than LyC; \citet{draine+2011}). So, the intricate radiative transfer of Ly$\alpha$ (for a review, see, e.g., \citet{dijkstra2017}) encodes complex information about the \ion{H}{1} that LyC photons must navigate to escape the galaxy. Though \ion{H}{1} governs the paths of Ly$\alpha$ and LyC photons, both are sensitive to significantly different column densities (e.g., for $\tau=1$, $\log_{10}(N_{\text{HI}}\,[\textrm{cm}^{-2}])\sim13.8$ and $\log_{10}(N_{\text{HI}}\,[\textrm{cm}^{-2}])\sim17$ for Ly$\alpha$ and LyC photons, respectively). The different column density sensitivities, coupled with the resonant nature of Ly$\alpha$, necessitates different interpretation approaches for Ly$\alpha$ and LyC signatures. LyC photons travel along a sightline until destroyed by intervening material (\ion{H}{1} or dust), but \ion{H}{1} can rescatter Ly$\alpha$ photons many times (provided dust does not destroy those Ly$\alpha$ photons). And crucially, unlike Ly$\alpha$ photons, LyC photons are not strongly sensitive to the kinematics of \ion{H}{1}. This means Ly$\alpha$ photons may not actually follow the observed sightlines, and could have originated from a different physical region than observed. So, Ly$\alpha$ properties do not necessarily correspond with the areas that they appear to come from, and may appear diffused to a much larger area than the progenitor \ion{H}{2} regions (e.g., \citet{ostlin+2014, hayes+2014, rivera-thorsen+2015, herenz+2016, bridge+2018, rasekh+2022, melinder+2023}).

Because of the complex connection between the regulation of Ly$\alpha$ and LyC photon escape, one of the Sunburst Arc's most remarkable features is its triple-peaked Ly$\alpha$ profile, which we investigate in this work. \citet{rivera-thorsen+2017} first reported this feature, originally predicted by \citet{behrens+2014}, and seldom observed in other galaxies \citep{izotov+2018d, vanzella+2018, 2020MNRAS.491.1093V}. %observed (not nearly as unambiguously) in another high redshift LyC leaker \citep{2020MNRAS.491.1093V}. 
As we present, with one exception, all observations of triple-peaked Ly$\alpha$ profiles in the Sunburst Arc seem associated with the LyC-leaking region. So, this unique Ly$\alpha$ profile may be connected to LyC escape. Observations and radiative transfer models of Ly$\alpha$ often explain this unique profile with a perforated covering shell of \ion{H}{1} that permits direct escape of Ly$\alpha$ and LyC photons through one or several holes, but otherwise scatters (attenuates) Ly$\alpha$ (LyC) photons \citep{behrens+2014, verhamme+2015, rivera-thorsen+2015, rivera-thorsen+2017}.

% Lead-in to rest of paper
In this work, we analyze rest-UV and rest-optical observations of the Sunburst Arc from slit spectroscopy with Magellan/MagE (\S\,\ref{ssec:obs_mage}) and archival multi-filter imaging with HST (\S\,\ref{ssec:obs_hst}). We specifically study the Ly$\alpha$ properties of regions of the galaxy and their $f_{\text{esc}}^{\text{LyC}}$ to build a physical picture of how Ly$\alpha$ and LyC photons escape the galaxy. To accomplish this, we fitted the Ly$\alpha$ spectra (\S\,\ref{ssec:methods_lya}) and calculated estimates of $f_{\rm{esc}}^{\rm{LyC}}$ in the spectroscopic apertures from the HST imaging (\S\,\ref{ssec:methods_fesc}). We compare common Ly$\alpha$ parameters to determine their mutual dependence and relation to LyC escape (\S\,\ref{sec:results}). In \S\,\ref{sec:disc} we present several hypotheses to explain the spatial variation of the Ly$\alpha$ velocity profiles and their connection to LyC escape, ultimately favoring unique source plane \ion{H}{1} morphologies as causative mechanisms for the observed Ly$\alpha$ and LyC signatures. We conclude in \S\,\ref{sec:con} with implications for future Ly$\alpha$ and LyC observations in lensed galaxies.

Throughout this work, we assume a flat $\Lambda$CDM cosmology with $H_0=70$ km s$^{-1}$ Mpc$^{-1}$, $\Omega_m=0.3$, and $\Omega_\Lambda=0.7$.

\section{Observations} \label{sec:obs}

This work used rest-frame Ly$\alpha$ spectra from Magellan/MagE (\S\,\ref{ssec:obs_mage}) as well as HST imaging of the rest-frame Ly$\alpha$ (HST/WFC3 F410M), LyC (HST/WFC3 F275W), near-UV continuum (HST/ACS F814W, WFC3 F390W, F555W, F606W) and rest-frame optical narrowband (HST/WFC3 F128N, F153M) (\S\,\ref{ssec:obs_hst}). See Tables \ref{tab:mage_log} and \ref{tab:hst_log} for descriptions of the respective Magellan and HST observations.

\subsection{Magellan spectroscopy}
\label{ssec:obs_mage}

%The foreground cluster strongly lenses the Sunburst Arc into four major arcs (Figure \ref{fig:arcs_mage}, left). We observed the Sunburst Arc at 9 locations (Figure \ref{fig:arcs_mage}, right) on the two largest arcs in the optical wavelengths using the Magellan Echellette (MagE) spectrograph \citep{2008SPIE.7014E..54M} mounted on the Magellan Baade Telescope of the Las Campanas Observatory in Chile. Of the 9 slits, 5 target images of a multiply-imaged LyC leaker in the galaxy, and the remaining 4 target combinations of multiple physically distinct regions of the galaxy that are not leaking LyC photons. The brightest image in the nonleaker slit M3 is an unusual, highly magnified object discussed by \citealt{vanzella+2020}, \citealt{diego+2022}, and \citealt{sharon+2022}, and concluded to be a single LBV star in outburst by \citealt{diego+2022}. 

\begin{figure*}[t!]
    \centering
    \includegraphics[width=\textwidth]{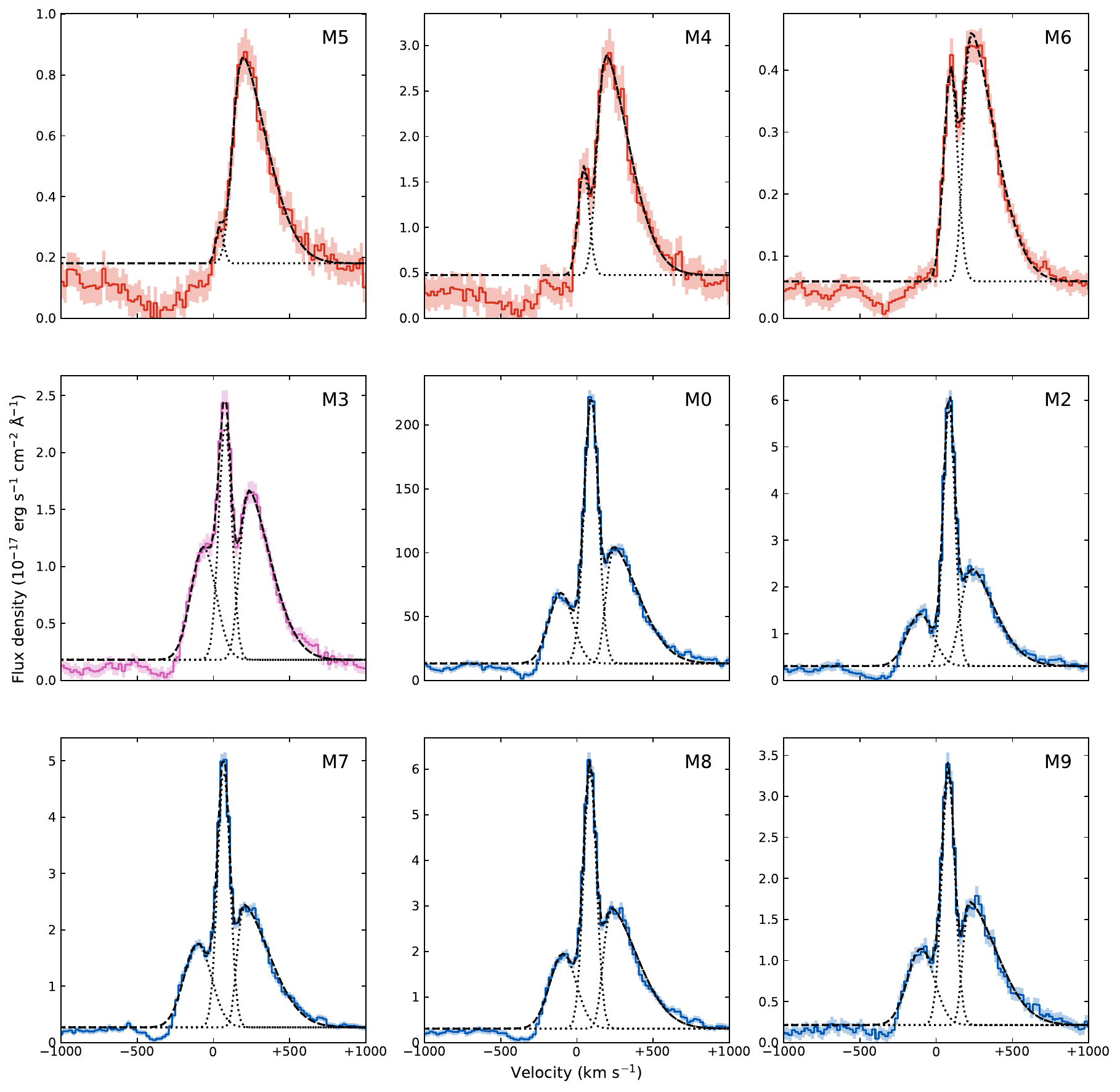}
    \caption{The magnification-corrected, rest-frame Ly$\alpha$ profiles, ordered from left to right, then top to bottom, first by increasing central Ly$\alpha$ peak strength relative to the redshifted Ly$\alpha$ peak in the non-LyC-leaking apertures, then the LyC-leaking apertures by the numeric order of their slit identifier. Those in red (top row) are non-LyC-leaking apertures (slits M5, M4, and M6), in pink (slit M3) is a non-LyC-leaking aperture targeting a highly magnified, exotic object (see \S\,\ref{sssec:disc_lya_m3} for more details), and in blue are the LyC-leaking apertures (slits M0, M2, M7, M8, and M9). The dashed black lines represent the overall fits and dotted black lines represent the individual Ly$\alpha$ peak fits, based on the median best-fit parameters from the fitting described in \S\,\ref{sssec:methods_lya_fwhm_vsep}. The shaded region represents the $2\sigma$ uncertainties of the flux densities. The observations clearly show a significant variety of Ly$\alpha$ profiles in the Sunburst Arc. Slit M0 has a much larger flux density scale due to poor observing conditions that prevented an accurate fluxing (see Table \ref{tab:mage_log} for more details about the observation).}
    \label{fig:lya_fits}
\end{figure*}

We observed 9 locations on the 2 largest arcs of the Sunburst Arc (Figure \ref{fig:arcs_mage}) in the observed-frame optical ($\sim3200-8500$ \AA) wavelengths with the Magellan Echellette (MagE) spectrograph \citep{2008SPIE.7014E..54M} mounted on the Magellan-I Baade Telescope (6.5m) at the Las Campanas Observatory in Chile. These spectra cover the rest-UV ($\sim950-2500$ \AA, including Ly$\alpha$) at the redshift of the Sunburst Arc. Apart from slit M0, which predated the acquisition of HST imaging, we compared the MagE slit viewing camera to HST images to precisely position the slits along the arc. M0 simply targeted the brightest part of the arc as measured in discovery imaging from the European Southern Observatory's New Technology Telescope \citep{2016A&A...590L...4D}. Of the 9 apertures, the HST imaging reveals that 5 apertures target images of the LyC-leaking region and 4 do not. One slit (M3) covers an unusual, highly magnified object previously discussed by \citet{vanzella+2020}, \citet{diego+2022}, \citet{sharon+2022}, \citet{pascale+2024}, and \citet{choe+2024} (image 4.8 in Figures \ref{fig:arcs_mage}, \ref{fig:lya_and_lyc_maps}), favored to be a supernova by \citet{vanzella+2020} and later a luminous blue variable (LBV) star in outburst by \citet{diego+2022}. \citet{sharon+2022} presented evidence the unusual source is not a transient event, and is likely due to an unusual lensing configuration, as suggested by \citet{diego+2022}. Table \ref{tab:mage_log} summarizes the MagE pointings. Rigby et al. (in prep.) will present the full observation log for each pointing. The MagE data were reduced following the same methods as described in \citet{2018AJ....155..104R}. Slits M0 and M3 and two conglomerate stacked spectra are the only previously published portions of these MagE observations (first published in \citet{rivera-thorsen+2017}, \citet{choe+2024}, and \citet{mainali+2022}, respectively). The reduction procedure of the new data presented here is identical to the reduction for slit M0 published in \citet{rivera-thorsen+2017}. The MagE observations of Ly$\alpha$ appear in Figure \ref{fig:lya_fits}.

\subsection{HST imaging}
\label{ssec:obs_hst}

We adopted some HST data used in this work as reduced and presented in other works: the ACS F814W and WFC3 F390W, F410M, F555W, and F606W observations in \citet{sharon+2022}, and the WFC3 F128N, F153M observations in \citet{kim+2023}. 

We analyzed new, ultra deep rest-LyC images, adding 26 additional orbits of F275W (GO-15949, PI: M. Gladders) to the 3 orbits of F275W previously presented in \citet{rivera-thorsen+2019} (GO-15418, PI: H. Dahle), resulting in a total of 29 orbits (86.9 ks) of integration time. The new F275W observations were structured in visits of typically three orbits, with two full-orbit integrations to minimize effects of charge transfer inefficiency, and two half-orbit integrations, providing a total of four frames to ensure good cosmic ray rejection and point spread function (PSF) reconstruction in each visit. Individual processed frames were taken from the Mikulski Archive for Space Telescopes and stacked using the tools in the \texttt{DrizzlePac} package. The final stacked image was astrometrically referenced to the F606W imaging, with a pixel scale of 0.03{\arcsec} pixel$^{-1}$ and a 0.6 pixel Gaussian drizzle drop size. The deeper exposures achieve a 5$\sigma$ depth of $m_{\rm{AB}}\approx29.9$ (compared to the corresponding 5$\sigma$ depth of $m_{\rm{AB}}\approx28$ from the shallower exposures in \citet{rivera-thorsen+2019}) but do not reveal any new LyC-leaking images or LyC morphology than presented in \citet{rivera-thorsen+2019}. Figure \ref{fig:lya_and_lyc_maps} shows the new LyC maps of the two main segments of the Sunburst Arc targeted by this work's MagE slit apertures. Table \ref{tab:hst_log} summarizes the HST observations used in this work.

All the HST data used in this paper can be found in MAST: \dataset[https://dx.doi.org/10.17909/t87g-a816]{https://dx.doi.org/10.17909/t87g-a816}.

%We used the HST/ACS F814W imaging from GO-15101 (PI: H. Dahle), previously presented by \citet{rivera-thorsen+2019}. See that work for details about the observations and data reduction. We combined two epochs of HST/WFC3 F275W observations from GO-15101 (PI: H. Dahle) and GO-15949 (PI: M. Gladders) to produce deeper observations. To do this, we...
\begin{deluxetable*}{rllllll}
    \tablecaption{HST observations \label{tab:hst_log}}
    \tablehead{
        \colhead{Camera} & \colhead{Filter} & \colhead{$\lambda_{\rm pivot}$} & \colhead{Width} & \colhead{$t_{\rm{exp}}$} & \colhead{Purpose} & \colhead{Program}
        \\
        & & [\AA] & [\AA] & [s] & &
    }
    \startdata
    ACS & WFC/F814W & 8333 & 2511 & 5280 & $f_{\rm{esc}}^{\rm{LyC}}$, SED fit & GO-15101
    \\
    \hline
    WFC3 & IR/F128N & 12832 & 159 & 16818 & SED fit & GO-15949
    \\
     & IR/F153M & 15322 & 685 & 5612 & SED fit & GO-15949
    \\
     & UVIS/F275W & 2710 & 405 & 5413 & $f_{\rm{esc}}^{\rm{LyC}}$ & GO-15418
    \\
     & UVIS/F275W & --- & --- & 81489 & $f_{\rm{esc}}^{\rm{LyC}}$ & GO-15949 
    \\
     & UVIS/F390W & 3924 & 894 & 3922 & Ly$\alpha$ off-band & GO-15949
    \\
     & UVIS/F410M & 4109 & 172 & 13285 & Ly$\alpha$ on-band & GO-15101
    \\
     & UVIS/F555W & 5308 & 1565 & 5616 & Ly$\alpha$ off-band, SED fit & GO-15101
    \\
     & UVIS/F606W & 5889 & 2189 & 5830 & SED fit & GO-15377
    \enddata

    \tablecomments{HST observations used in this work. From left to right: HST instrument, filter, pivot wavelength of filter in angstroms, filter width in angstroms, exposure time in seconds, purpose for this work, and program ID.}
\end{deluxetable*}

%GO-15101 (PI: H. Dahle) also observed the Sunburst Arc with the HST/WFC3 F410M filter, which captures Ly$\alpha$ photons at the arc's redshift. As described in \S\,\ref{ssec:methods_nb_lya}, we used this and continuum observations in F390W and F555W from GO-15101 and GO-15949 to create a narrowband Ly$\alpha$ emission map of the arc.

\section{Data Analysis}

% Introduce the MC measurement process

\subsection{Redshifts}

%Redshifts...

We adopted the spectroscopic redshifts presented by \citet{mainali+2022} (Table 1 therein), measured from the narrow component of the strong rest-optical nebular emission line [\ion{O}{3}] 5007 {\AA} of the MagE targets, captured with the Folded-port InfraRed Echellette (FIRE) \citep{simcoe+2013} mounted on the Magellan-I Baade Telescope. See Figure 1 in \citet{mainali+2022} for a comparison between the FIRE and MagE pointings (all the MagE pointings discussed in this paper have overlapping FIRE pointings) and their Section 3.3 for a description of how they computed the redshifts of the targeted regions of the galaxy. We did not adopt a single redshift for the galaxy because it is critical to accurately place each MagE spectrum into the rest frame while, e.g., stacking the spectra, or accurately determining peculiar velocities.

\subsection{Stacking spectra}

\begin{figure}[ht]
    \centering
    \includegraphics[width=0.7\columnwidth]{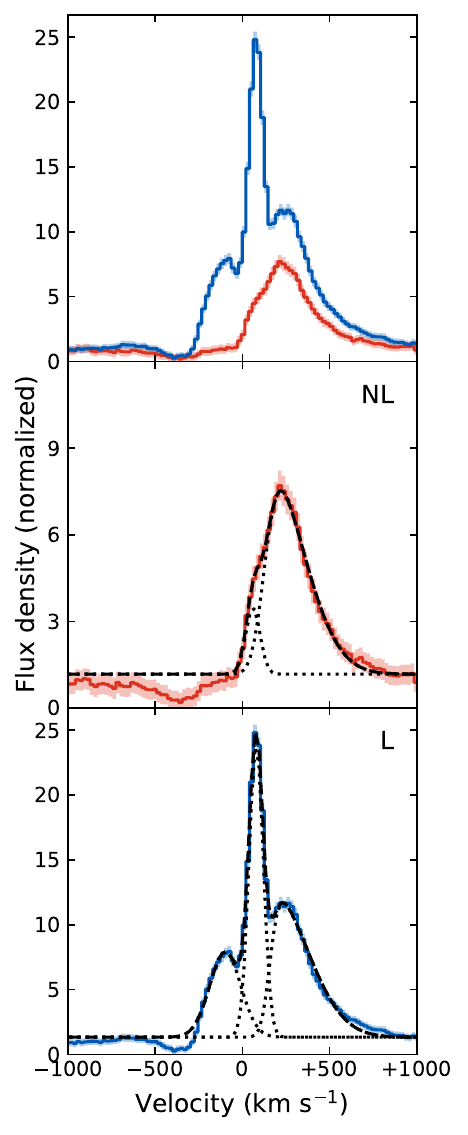}
    \caption{Ly$\alpha$ profiles of the stacked LyC-leaking and non-LyC-leaking MagE slit apertures with fits. \textit{Top}: Ly$\alpha$ profiles of the stacked LyC leaker (above in blue) and non-LyC-leaking regions (below in red). The shaded regions indicate the 3$\sigma$ uncertainty of the flux density. \textit{Center}: Stacked Ly$\alpha$ profile of the non-LyC-leaking apertures with the total fit in dashed black and individual peak fits in dotted black. \textit{Bottom}: Stacked LyC leaker Ly$\alpha$ profile with the total fit in dashed black and individual peak fits in dotted black.}
    \label{fig:lyastack}
\end{figure}

\citet{rivera-thorsen+2019} identified the LyC-leaking images in the Sunburst Arc as a compact, star-forming region, which \citet{mainali+2022} used to justify stacking the MagE spectra of the LyC-leaking and non-LyC-leaking images to compare the two sets of regions. We repeated this stacking (labeling the stacked LyC-leaking and non-LyC-leaking spectra as L and NL, respectively), but excluded slit M0 from the LyC-leaking stack due to the poor observing conditions (see Table \ref{tab:mage_log}) that prevented an accurate fluxing, and the presence of a foreground galaxy in the aperture (see the diffuse emission southeast of image 1.5 in Figure \ref{fig:arcs_mage}), and also excluded slit M3 from either stack, as it includes a bright image of an extremely magnified area of the galaxy \citep{diego+2022, sharon+2022} that may not be representative of the broader region it is embedded in (see \S\,\ref{sssec:disc_lya_m3} for more details). To create the stacks, we normalized the individual spectra used to create a stack by their median flux density between $1267-1276$ {\AA} in the rest frame, interpolated them to a common set of identically-sized wavelength bins, and then averaged their flux densities at each bin. Figure \ref{fig:lyastack} shows the stacked Ly$\alpha$ spectra.

\subsection{Creating narrowband Ly\texorpdfstring{$\alpha$}{man-alpha} maps}
\label{ssec:methods_nb_lya}

To supplement the spectroscopic Ly$\alpha$ data with spatial Ly$\alpha$ information, we created Ly$\alpha$ images by estimating the contribution of Ly$\alpha$ to the emission observed in the F410M filter using an approach similar to that described in Section 2.5 of \citet{kim+2023}. We fit SED models to the integrated F153M, F128N, F814W, F606W, and F555W photometry of four images of the multiply-imaged source galaxy, which include flux from image families 3, 4, 8, 9, and 10 \citep[using the nomenclature of][]{sharon+2022}. These four fits should produce nearly identical SEDs because they are fit to different lensed images of the same source. Fitting four different SEDs provides a direct estimate of any systematic variations caused by differential magnification between the different emission clumps in the lensed galaxy.  

We used \textsc{prospector}, an MCMC-based stellar population synthesis and parameter inference framework \citep{Conroy_2010, emceehammer, johnson+2021} 
% (\citealp{Conroy_2010}; \citealp{emceehammer}; \citealp{johnson21})
to fit SED models with a delayed-$\tau$ (with age and $e$-folding time of SFR as free parameters) + constant star formation model, and also fitted for stellar mass, dust attenuation (using the Calzetti attenuation model), and the gas ionization parameter log $U$. We used a fixed value for stellar metallicity ($\rm{log}(Z/Z_{\odot}) = -0.33$, informed by spectroscopy), and a Kroupa initial mass function \citep{kroupa2001}, and evaluated nebular continuum and line emission via the CLOUDY component of the embedded Flexible Stellar Population Synthesis (FSPS; \citealt{ben_johnson_2022_7113363}) framework (see \cite{byler2017} for details). We took the stellar and nebular continuum from the highest posterior probability SED model as a proxy for the true continuum. The SEDs and filter transmission curves (except F606W and F814W) appear in Figure 7 of \citet{kim+2023}.
%We took the stellar and nebular continuum components of the maximum a posteriori SED model as a proxy for the continuum emission observed from each image.
%Since the images have different brightnesses owing to variations in magnification, we normalized each SED at an observed wavelength of 1 $\mu$m.
We corrected the SEDs for Milky Way dust extinction according to the 3-dimensional interstellar dust extinction map of \citet{green+2015}.

\begin{figure*}
    \centering
    \includegraphics[width=0.9\textwidth]{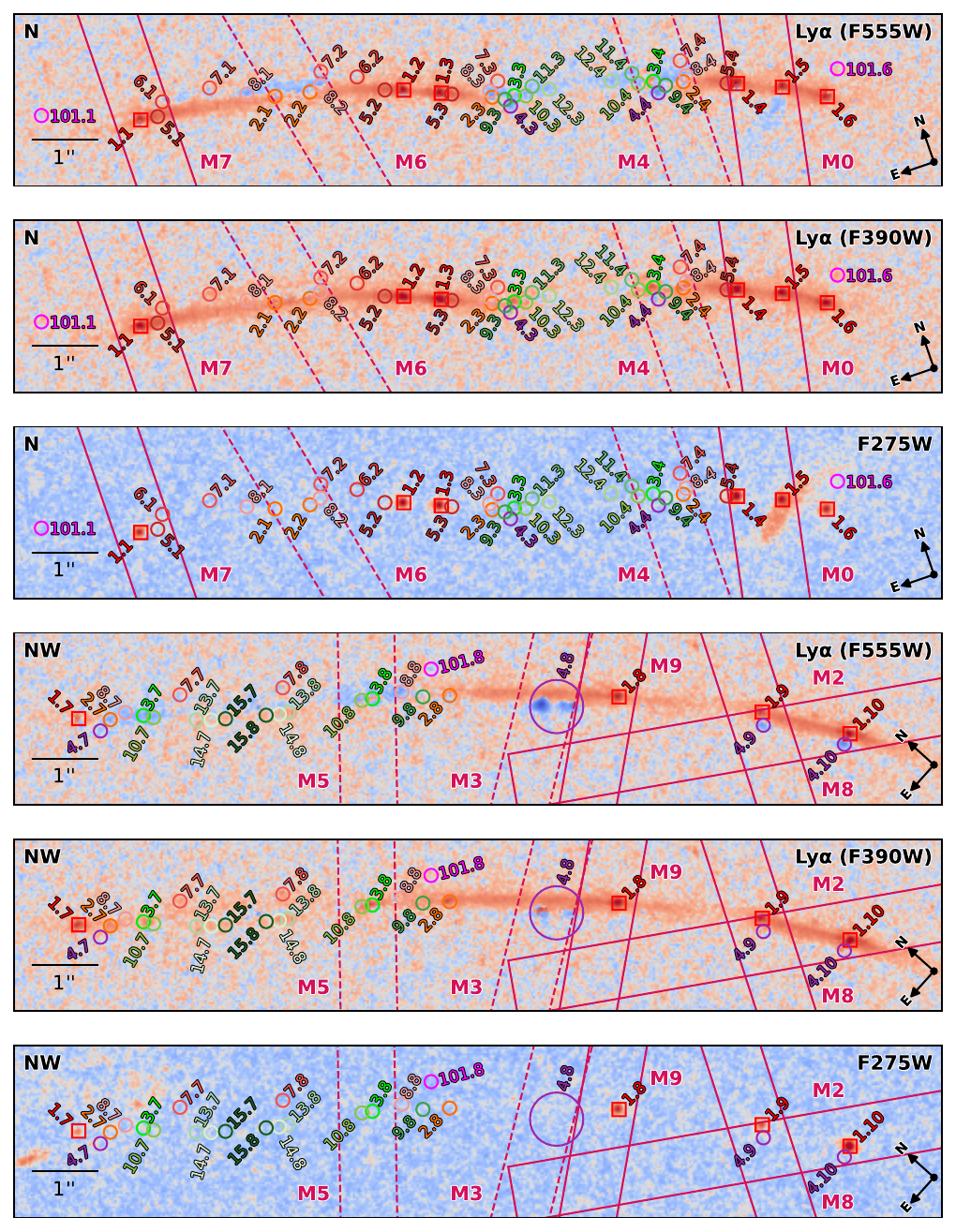}
    \caption{The rest-LyC (F275W) and Ly$\alpha$ maps of the north (N) and northwest (NW) segments of the Sunburst Arc, with the MagE slit apertures overlaid in magenta and the identified images of the Sunburst Arc marked and labeled following the color and naming scheme of \citet{sharon+2022}. Boxes mark images of the LyC-leaking region and circles mark all other images. The filter in parentheses in the Ly$\alpha$ map labels indicates the filter used to estimate the background continuum contribution to the count rate of F410M, which includes the Ly$\alpha$ line.}
    \label{fig:lya_and_lyc_maps}
\end{figure*}

These SEDs did not model attenuation from the Ly$\alpha$ forest, which may cause an over-subtraction of the continuum. To test if this was an issue, we normalized each SED at an observed wavelength of 1 $\mu$m and compared the SED count rates in the on-band (on-Ly$\alpha$) F410M filter with the count rates predicted for F410M from the stacked MagE LyC-leaking and non-LyC-leaking spectra after masking out Ly$\alpha$ and other emission lines, and convolving the data with a boxcar kernel. We found no significant difference between the predicted F410M continuum count rates from the SED fitting and the Ly$\alpha$-removed, scaled MagE spectra. %We found no significant difference between the count rates predicted by the SED fitting and the Ly$\alpha$-removed, scaled MagE spectra, as expected.
%Table X summarizes the results, which demonstrated that the count rates between the MagE spectra and SEDs were not significantly different, dispelling any concern about strong Ly$\alpha$ forest effects.

We calculated the Ly$\alpha$ emission as Ly$\alpha$ flux equal to on-band flux minus off-band flux $\times$ a scaling factor. We estimated the scaling factor as the mean of the individual scaling factors of the different continuum models described above. We used the standard deviation of the individual scaling factors of the models as an estimate of the systematic uncertainty in the scaling factor. We estimated the statistical uncertainty in the continuum subtraction from regions of empty sky in the data. We made two maps: one with F390W as the off-band filter---which samples continuum closer to Ly$\alpha$ but includes Ly$\alpha$ and Ly$\alpha$ forest-attenuated continuum---and one with F555W as the off-band filter---which samples continuum farther from Ly$\alpha$. Figure \ref{fig:lya_and_lyc_maps} shows the Ly$\alpha$ maps of the two segments of the Sunburst Arc for both off-band filter choices. We report uncertainties of measurements made from the Ly$\alpha$ maps with the combined systematic and statistical uncertainties.

\subsection{Measuring \texorpdfstring{$f_{\text{esc}}^{\text{LyC}}$}{the LyC escape fraction}} \label{ssec:methods_fesc}

\citet{rivera-thorsen+2019} previously computed $f_{\text{esc}}^{\text{LyC}}$ for the images of the LyC-leaking region, and more recently, \citet{pascale+2023} have indirectly estimated the LyC-leaking region's $f_{\text{esc}}^{\text{LyC}}$ from the observed nebular continuum and nebular line strengths. The MagE apertures include 5 slits that cover 6 of the LyC-leaking images investigated in \citet{rivera-thorsen+2019}, but also 4 slits that only cover images of the galaxy that do not leak LyC (Figure \ref{fig:arcs_mage}). We computed the IGM-unattenuated, absolute LyC escape fraction $f_{\text{esc}}^{\text{LyC}}$ within each MagE aperture. We adapted the method \citet{rivera-thorsen+2019} used to calculate the apparent, relative LyC escape fraction, summarized in their Equation S3. Briefly, they used the non-ionizing rest-UV continuum F814W observations (unaffected by \ion{H}{1} absorption) in tandem with theoretical, intrinsic Starburst99 \citep{leitherer+1999} spectra to compute the expected flux in F275W if the sightline were completely transparent to LyC. %They used 4 pixel wide apertures centered on the numbered squares in Figure \ref{fig:arcs_mage} to determine the flux in the F275W and F814W filters. 
To instead compute the IGM-unattenuated, absolute LyC escape fractions, we shifted the IGM transmission factor in their Equation S3 to the opposite side of the equation and used dust-extincted Starburst99 spectra instead of the intrinsic Starburst99 spectra.

\begin{figure}
    \centering
    \includegraphics[width=\columnwidth]{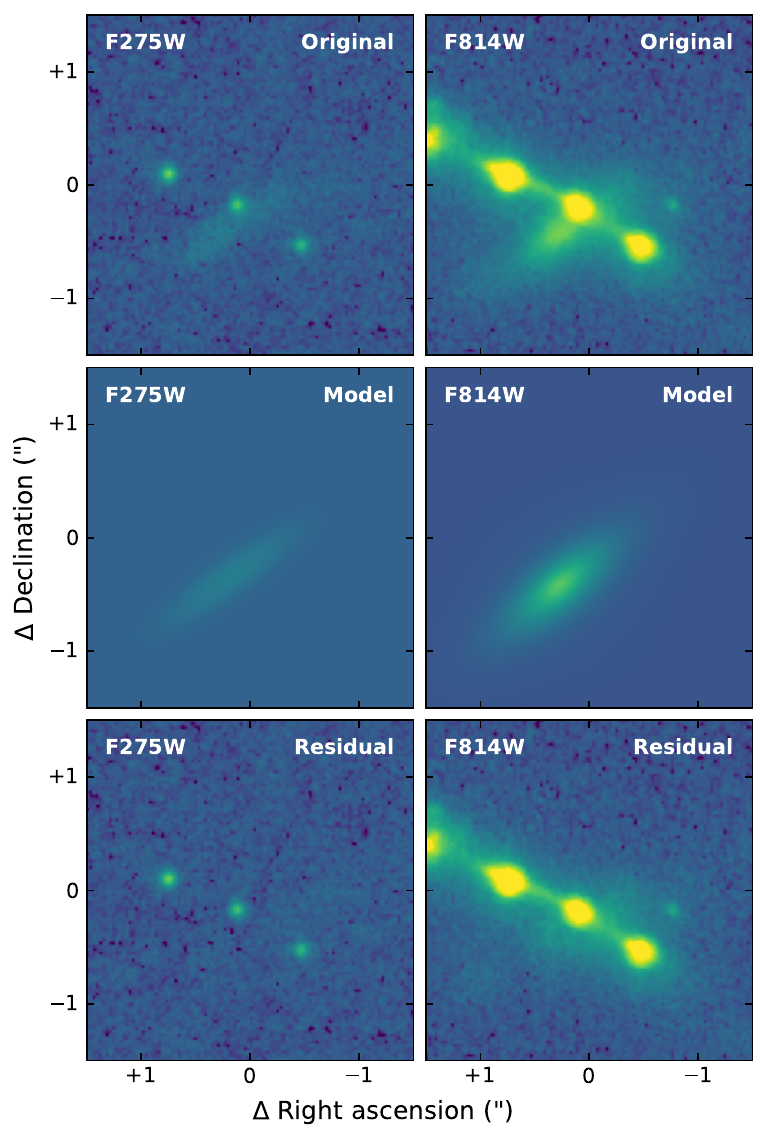}
    \caption{Results of using \textsc{Galfit} to remove a foreground galaxy over an image of the LyC leaker (image 1.5 in Figure \ref{fig:arcs_mage}) from the F275W and F814W images. Each column uses the same color map with a cube root scale, with minima (maxima) of the masked median of the cutout minus $2$ $\times$ (plus $100$ $\times$) the standard deviation of the source-masked cutout. The \textsc{Galfit} models have a flat background of 0.}
    \label{fig:galfit_result}
\end{figure}

To prepare the F275W and F814W images to measure the necessary photometry, we (1) removed a foreground interloping galaxy, and (2) subtracted the background level from the images. The foreground galaxy is bright in F275W, contaminating slit M0 (see the diffuse emission SE of image 1.5 in Figure \ref{fig:arcs_mage}). We subtracted the galaxy from the F275W and F814W imaging with the galaxy surface brightness profile modeling software \textsc{Galfit} \citep{peng+2010}. We fitted the galaxy with a single Sérsic profile in both F275W and F814W. We computed PSF models for both bands using in-field stars. The \textsc{Galfit} results appear in Figure \ref{fig:galfit_result}.

The summed F275W flux in a MagE aperture is sensitive to the background level of the drizzled image (and thus so is $f_{\text{esc}}^{\text{LyC}}$). Background level overestimates will underpredict $f_{\text{esc}}^{\text{LyC}}$, while underestimates will overpredict $f_{\text{esc}}^{\text{LyC}}$. To accurately model the complex background level, we used the Python Photutils package \citep{photutils1-5-0}. We iterated the background modeling parameters until the F275W flux in the non-LyC-leaking apertures was consistent with minimal flux. 

Although these steps make the images ready for photometry measurements, unlike \citet{rivera-thorsen+2019}, this work aims to closely compare LyC escape fractions to spectroscopic properties. The LyC escape fractions are measurable on a much smaller scale than the spectroscopic properties, since the latter are associated with a slit aperture (to compare, \citet{rivera-thorsen+2019} used square apertures $\sim0.1\arcsec$ across to measure photometries of images, as opposed to the $0.85\arcsec$ or $1\arcsec$ width of the MagE slit apertures). For this reason, it is important to match the LyC escape fraction measurements to the same physical scales sampled by the ground-based spectra for the most direct comparison between the spectroscopic and photometric properties. To accomplish this, we measured a single LyC escape fraction measurement for each slit aperture by first convolving the background-subtracted HST imaging with a 2-dimensional Gaussian kernel with a full width at half maximum (FWHM) matching the exposure time-weighted average of the observation's seeing conditions (Table \ref{tab:mage_log}), accounting for the airmass of each exposure of each slit aperture (Rigby et al. in prep.). We then measured the F275W and F814W fluxes associated with the LyC escape fraction measurement as the total flux in either filter inside the intersection of masks of the slit aperture and the Sunburst Arc, proceeding with the measured photometry in otherwise the same fashion as \citet{rivera-thorsen+2019}. See Table \ref{tab:f_esc} for the results.
%Following this, we convolved the background-substracted HST imaging with a Gaussian PSF with a full width at half maximum (FWHM) matching the MagE seeing conditions (Table \ref{tab:mage_log})---as recorded from the guide star during observations---to make our comparison to the MagE spectroscopy fair. Since each spectrum had multiple pointings, we determined the overall PSF from the exposure time-weighted average of each pointing's seeing conditions. 
%If an aperture's observations spanned multiple nights, we used an exposure time-weighted average of each observation's seeing conditions as the overall PSF.

For each slit aperture and each filter, we estimated the uncertainty of each pixel in the images as the standard deviation of the flux densities outside of the arc mask but inside the aperture. We conservatively assumed a constant 10\% uncertainty for the Starburst99 stellar continuum fits of each spectrum. We propagated these uncertainties through the calculation of the LyC escape fractions.

\input{f_esc_lyc_measurements_table.tex}

%\begin{deluxetable}{cccc}[t]

%\tablecaption{HST-based properties of the MagE spectra \label{tab:f_esc}}
%\tablehead{
%    \colhead{Slit} & \colhead{$F_{275}$} & \colhead{$F_{814}$} & \colhead{$f_{\rm{esc}}^{\rm{LyC}}$}
%}

%\startdata
%M5 & $4.1\pm0.1$ & $215\pm1$ & $2.3\pm0.8$
%\\
%M4 & $-1.8\pm0.1$ & $374.9\pm0.4$ & $-0.6\pm0.2$
%\\
%M6 & $10.1\pm0.2$ & $379.7\pm0.8$ & $3\pm1$
%\\
%M3 & $7.5\pm0.2$ & $393.0\pm0.6$ & $2.3\pm0.8$
%\\
%\hline
%M0 & $70.4\pm0.1$ & $492.2\pm0.8$ & $17\pm6$
%\\
%M2 & $89.0\pm0.2$ & $594\pm1$ & $18\pm7$
%\\
%M7 & $31.1\pm0.2$ & $297.9\pm0.2$ & $12\pm5$
%\\
%M8 & $23.6\pm0.2$ & $186.9\pm0.6$ & $15\pm6$
%\\
%M9 & $34.5\pm0.2$ & $291\pm2$ & $14\pm5$
%\enddata

%\tablecomments{From left to right: slit label, flux density in the HST/WFC3 F275W and HST/ACS F814W filters ($10^{-20}$ erg s$^{-1}$ cm$^{-2}$ {\AA}$^{-1}$), and $f_{\text{esc}}^{\text{LyC}}$ (\%), all computed according to \S\,\ref{ssec:methods_fesc}.}

%\end{deluxetable}

\subsection{Measuring Ly\texorpdfstring{$\alpha$}{man-alpha} properties}
\label{ssec:methods_lya}

For all the following quantities, we used a Monte Carlo (MC) measurement process of 10,000 iterations with a burn-in simulation of 1,000 iterations to estimate their values, in which we assumed the flux density and flux density uncertainty associated with each wavelength bin of the MagE spectra corresponded respectively to the mean and standard deviation of a Gaussian distribution. In each iteration, we drew a random sample from the Gaussian distributions corresponding to each wavelength bin to create a new mock spectrum that we then used to measure the discussed quantities. In the main text, figures, and tables, for a given measurement's distribution, we cite the median $M$, and absolute differences between $M$ and the 16th and 84th percentiles $A$ and $B$, respectively, of the Monte Carlo simulation results as $\rm{M}_{-\rm{A}}^{+\rm{B}}$.

In all Ly$\alpha$ profiles, we measured the FWHM of the central and redshifted Ly$\alpha$ peaks (\S\,\ref{sssec:methods_lya_fwhm_vsep}), the ratio $f_{\rm{min}}/f_{\rm{cont}}$ (introduced in \S\,\ref{sssec:methods_lya_fmin_fcont}), the rest-frame Ly$\alpha$ equivalent width (EW) (\S\,\ref{sssec:methods_lya_ew}), and central escape fraction $f_{\rm{cen}}$ (\citet{naidu/matthee+22}; introduced later in \S\,\ref{sssec:f_cen}). If the Ly$\alpha$ profile also had a blueshifted Ly$\alpha$ peak, we measured its FWHM and the velocity separation between the redshifted and blueshifted Ly$\alpha$ peaks (\S\,\ref{sssec:methods_lya_fwhm_vsep}). In non-stacked spectra, we also measured the magnification-corrected Ly$\alpha$ luminosity (\S\,\ref{sssec:methods_luminosity}). Table \ref{tab:lya_params} contains the measurements.

%In all Ly$\alpha$ profiles, we measured the equivalent width (EW$_{\text{Ly}\alpha}$), luminosity, central escape fraction $f_{\text{cen}}$ (\citealt{naidu/matthee+22}; introduced later in Section \ref{sssec:f_cen}), and FWHM of the redshifted peak. If the profile had a central peak, we also measured the FWHM of this peak. If the profile also had a blueshifted peak, we measured its FWHM, the separation between the red and blueshifted peaks, and the ratio $f_{\text{min}}/f_{\text{cont}}$ of the `minimum' (really the maximum, as we discuss later) flux density between the red and blueshifted peaks and the local continuum flux density.

\subsubsection{Equivalent width}
\label{sssec:methods_lya_ew}

The equivalent width of an absorption or emission line is a measure of its strength relative to the continuum level. In this work, we choose the convention that emission lines have positive equivalent widths, and vice versa for absorption lines, so that the equivalent width is
\begin{equation}
    \text{EW} = -\int_{\lambda_1}^{\lambda_2}\left(1-\frac{F_\lambda}{F_c}\right)\,d\lambda.
    \label{eq:ew}
\end{equation}
Here $F_\lambda$ is the flux density, $F_c$ the continuum flux density, and $\lambda_1, \lambda_2$ are the integration bounds over the spectral feature.

We applied Equation \ref{eq:ew} to compute the rest-frame Ly$\alpha$ EW by integrating between $1212-1221$ {\AA} in the rest frame, as in \citet{2017ApJ...844..171Y}. We sampled the continuum flux density as the median flux density between $1221-1225$ {\AA} in the rest frame.

%The leaking spectra show a distinctly stronger Ly$\alpha$ feature than the nonleaking regions (excluding the aforementioned M3), with EWs $\sim24-33$\,{\AA}, as compared to EWs $\sim7-15$\,{\AA} in the nonleakers.

\subsubsection{Luminosity}
\label{sssec:methods_luminosity}

To compute the Ly$\alpha$ luminosity, we integrated the continuum-subtracted (taking the local continuum as in $\S$\,\ref{sssec:methods_lya_ew}), magnification-corrected flux density between $1212-1221$ {\AA} in the rest frame.% We assumed a flat $\Lambda$CDM cosmology with $H_0=70$\,km\,s$^{-1}$\,Mpc$^{-1}$, $\Omega_{0,m}=0.3$, $\Omega_{0,\Lambda}=0.7$.

%We found $\text{log}(L_{\text{Ly}\alpha}(\text{erg s}^{-1}))\sim41.4-42.2$ in the nonleaker and $\sim42.3-42.5$ in the leaker.

\subsubsection{\texorpdfstring{$f_{\text{cen}}$}{Central escape fraction}}
\label{sssec:f_cen}

The central fraction of Ly$\alpha$ flux, $f_{\rm{cen}}$, introduced by \citet{naidu/matthee+22}, is
\begin{equation}
    f_{\text{cen}}=\frac{\text{Ly$\alpha$ flux between $\pm$100\,km s$^{-1}$}}{\text{Ly$\alpha$ flux between $\pm$1000\,km s$^{-1}$}},
\end{equation}
so named because it integrates the flux densities in the specified velocity bands centered on the wavelength of Ly$\alpha$. The numerator, which targets a narrow band about the Ly$\alpha$ wavelength, is sensitive to Ly$\alpha$ photons that have not been rescattered (attenuated) much by \ion{H}{1} (dust). But due to the `random walk' nature of a Ly$\alpha$ photon's radiative transfer, the numerator also includes any Ly$\alpha$ photons that randomly walk back to the central wavelength band after significant reprocessing. If significantly underdense sightlines exist to the areas of Ly$\alpha$ production, Ly$\alpha$ photons escaping through them should appear in the central wavelength band. The denominator captures virtually all Ly$\alpha$ flux. Thus, $f_{\rm{cen}}$ represents the relative strength of minimally scattered Ly$\alpha$ photons compared to the total number of Ly$\alpha$ photons. \citet{naidu/matthee+22} predict that $f_{\rm{cen}}$ should correlate with the LyC escape fraction, since LyC photons must navigate similar obstacles (i.e., \ion{H}{1} and dust) to escape a galaxy. 

%We measured this quantity in all the MagE spectra and found strong differences between the leaker ($\sim30\%$) and the nonleaker ($\sim4-20\%$), in line with the picture that the leaker's underdense channels permitting LyC escape also allow more minimally scattered Ly$\alpha$ photons to escape.

\subsubsection{Peak widths and separation}
\label{sssec:methods_lya_fwhm_vsep}

%The remaining measurement process is contingent upon the structure of the Ly$\alpha$ profile in question. We identified three cases: (i) there is a redshifted, blueshifted, and central peak, (ii) there is only a redshifted and central peak, and (iii) there is only a redshifted peak. Most (6/9) of the MagE spectra fall into (i); only two (M4 and M6) fall into case (ii) and only M5 into case (iii). In case (i) and (ii), we additionally measured the FWHM of the central peak. In case (i), we also measured the FWHM of the blueshifted peak, the separation between the blue and red peaks, and the ratio $f_{\text{min}}/f_{\text{cont}}$ between the `minimum' flux density between the peaks and the local continuum flux density.

Determining the width and separation of the Ly$\alpha$ peaks depends on the structure of the Ly$\alpha$ profile in question. We organized the spectra into two cases: (i) there is a redshifted, blueshifted, and central Ly$\alpha$ peak, and (ii) there is not a clear blueshifted Ly$\alpha$ peak. This dichotomy implies that all the spectra have central Ly$\alpha$ peaks, even though slit M5 does not clearly show a central Ly$\alpha$ peak. We still include slit M5 in the two-case dichotomy (the latter case) in order to constrain the possibility of a faint, unresolved central Ly$\alpha$ peak. Most (all the LyC-leaking spectra and slit M3) of the spectra fall into case (i), and only some of the non-LyC-leaking spectra occupy case (ii) (slits M4, M5, and M6). %To directly measure the FWHM of each peak and the separation of the red and blue peaks $v_{\rm{sep}}$, we fitted each spectrum with a combination of Gaussian and skewed Gaussian functions.

\citet{rivera-thorsen+2017} previously modeled the central peak as a Gaussian function of the form
\begin{equation}
    \Gamma_\lambda(v) = \alpha e^{-\frac{1}{2}((v-\mu)/\sigma)^2},
\end{equation}
where $\alpha$ is its amplitude, $\mu$ its centroid, and $\sigma$ its standard deviation. Other authors (e.g., \citealt{mallery+2012}, \citealt{u+2015}, \citealt{cao+2020}) have treated double-peaked Ly$\alpha$ profiles (those with a redshifted and blueshifted Ly$\alpha$ peak) as two skewed Gaussian functions that follow the form
\begin{equation}
    \xi_\lambda(v) = \Gamma_\lambda(v)\left[1+{\rm{erf}}\left(\frac{\omega(v - \mu)}{\sqrt{2}\sigma}\right)\right].
\end{equation}
Here $\omega$ is the skewness, which controls how skewed the distribution is, and ${\rm{erf}}(x)$ is the error function, a complex function with complex variable $x$ defined as
\begin{equation}
    {\rm{erf}}(x)=\frac{2}{\sqrt{\pi}}\int_0^xe^{-t^2}\,dt.
\end{equation}
Right-skewed distributions (i.e., a redshifted Ly$\alpha$ peak) have $\omega>0$ and left-skewed distributions (i.e., a blueshifted Ly$\alpha$ peak) have $\omega<0$.

We combined these two approaches by simultaneously fitting the Ly$\alpha$ profiles to a combination of Gaussian and skewed Gaussian functions with the $\texttt{curve\_fit()}$ function in the SciPy Python package. We directly measured each peak's FWHM and the separation between the redshifted and blueshifted Ly$\alpha$ peaks $v_{\rm{sep}}$ from the resulting fit. To determine the intrinsic FWHM of a peak, we assumed the overall FWHM measured from the observed spectrum as Gaussian, and deconvolved it with the assumed Gaussian line spread function of the instrument (${\rm FWHM}=c/R$), randomly sampling the spectral resolution at each iteration of the MC procedure from a Gaussian distribution described by the spectral resolutions in Table \ref{tab:mage_log}. We determined the spectral resolution of the stacked spectra by averaging the spectral resolutions of their composite spectra and adding the associated uncertainties in quadrature. The resulting FWHM is what we report in Table \ref{tab:lya_params}. We fitted the following functions for the two aforementioned cases:
\begin{itemize}
    \item[(i)] $\xi_\lambda^{\rm{blue}}+\xi_\lambda^{\rm{red}}+\Gamma_\lambda+c$, and
    \item[(ii)] $\xi_\lambda^{\rm{red}}+\Gamma_\lambda+c$,
\end{itemize}
where $c$ is a scalar continuum contribution. In case (ii), even when there is not a clear central Ly$\alpha$ peak resolved from the redshifted Ly$\alpha$ peak, attempting to fit a central Gaussian component can directly constrain the strength of any unresolved central Ly$\alpha$ peak. The stacked spectrum of the non-LyC-leaking apertures is strongly suggestive of such an instance, since it shows a noticeable bulge on the blueward side of the redshifted Ly$\alpha$ peak poorly reproduced by a single skewed Gaussian (Figure \ref{fig:lyastack}). The best-fit model parameters appear in Table \ref{tab:fit_results}.

%In case (i) we fit two skewed Gaussian functions (the red and blue peaks) and a Gaussian function (the central peak) to the profile. In case (ii) we fit a skewed Gaussian function (the red peak) and a Gaussian function (the central peak). In case (i)

%In case (i), we measured the FWHM of the blueshifted peak, the separation between the blue and red peaks, and the ratio $f_{\text{min}}/f_{\text{cont}}$ between the `minimum' flux density between the peaks and the local continuum flux density.

%In case (i) and (ii), following \citep{rivera-thorsen+2017}, we continued by fitting a narrow band of the central peak as a Gaussian profile, setting the central peak's FWHM and, as we discuss next, $f_{\text{min}}/f_{\text{cont}}$. We then subtracted this fit from the overall profile to isolate the red and blue peaks and measured their FWHMs and separation. In case (iii), we made no additional modifications to the spectrum before measuring the red peak's FWHM.

\subsubsection{\texorpdfstring{$f_{\text{min}}/f_{\text{cont}}$}{Trough-to-continuum ratio}}
\label{sssec:methods_lya_fmin_fcont}

Instead of a triple-peaked profile, Ly$\alpha$ profiles are almost always single- or double-peaked, showing some combination of a redshifted and blueshifted Ly$\alpha$ peak, but no clear, central Ly$\alpha$ peak of directly escaping Ly$\alpha$ photons. In this case, a common parameter closely connected to the \ion{H}{1} scattering environment is the ratio between the minimum flux density between the redshifted and blueshifted Ly$\alpha$ peaks, $f_{\text{min}}$, and the flux density of the local continuum, $f_{\text{cont}}$ (e.g., \citet{jaskot+2019}). Functionally, this quantity directly constrains the strength of a central peak of direct escape Ly$\alpha$ photons. Usually, such a peak (if it exists) is completely unresolved due to much stronger, nearby redshifted and blueshifted Ly$\alpha$ peaks. Clearly this is not the case in the Sunburst Arc, so this measurement breaks down, as often the peak Ly$\alpha$ intensity appears between the redshifted and blueshifted Ly$\alpha$ peaks (Figures \ref{fig:lya_fits}, \ref{fig:lyastack}). 

Instead, we measured the central Ly$\alpha$ peak's fitted amplitude as $f_{\text{min}}$, as this interpretation captures the `spirit' of $f_{\text{min}}/f_{\text{cont}}$; to gauge the prevalence of direct escape Ly$\alpha$ photons (which constitute the central Ly$\alpha$ peak) and probe the \ion{H}{1} column density. We took $f_{\text{cont}}$ as the local continuum flux density fitted in the functions described in \S\,\ref{sssec:methods_lya_fwhm_vsep}.

\input{lya_measurements_table.tex}

\section{Results} \label{sec:results}

Previous work has investigated a wide range of correlations between Ly$\alpha$ and LyC parameters, which is critical to understand the intimate relation between Ly$\alpha$ and LyC escape. Observations have found positive correlations between:
\begin{itemize}
    \item $f_{\rm{esc}}^{\rm{LyC}}-\rm{Ly\alpha \ EW}$ \citep{verhamme+2017, fletcher+2019, flury+2022b, saldana-lopez+2022, pahl+2023}, 
    \item $f_{\rm{esc}}^{\rm{LyC}}-\rm{Ly\alpha \ FWHM}$ \citep{kramarenko+2023},
    \item $f_{\rm{esc}}^{\rm{LyC}}-f_{\rm{cen}}$ \citep{naidu/matthee+22}, 
    \item $f_{\rm{esc}}^{\rm{LyC}}-f_{\rm{min}}/f_{\rm{cont}}$ \citep{gazagnes+2020}, 
    \item $v_{\rm{sep}}-\rm{Ly\alpha \ FWHM}$ \citep{verhamme+2018, kerutt+2022},
    \item and $f_{\rm{min}}/f_{\rm{cont}}-\rm{Ly\alpha \ EW}$ \citep{jaskot+2019},
\end{itemize}
negative correlations between:
\begin{itemize}
    \item $f_{\rm{esc}}^{\rm{LyC}}-v_{\rm{sep}}$ \citep{verhamme+2017, izotov+2018d, izotov+2021, izotov+2022, gazagnes+2020, flury+2022b, naidu/matthee+22},
    \item $f_{\rm{min}}/f_{\rm{cont}}-v_{\rm{sep}}$ \citep{jaskot+2019},
    \item $v_{\rm{sep}}-\rm{Ly\alpha \ EW}$ \citep{verhamme+2017, jaskot+2019, marques-chaves+2020a},
    \item and $\rm{Ly\alpha \ FWHM}-\rm{Ly\alpha \ EW}$ \citep{hashimoto+2017}, 
\end{itemize}
and noncorrelations between:
\begin{itemize}
    \item $f_{\rm{esc}}^{\rm{LyC}}-\rm{Ly\alpha \ EW}$ \citep{mestric+2020}, 
    \item $v_{\rm{sep}}-\rm{Ly\alpha \ EW}$ \citep{kerutt+2022},
    \item and $\rm{Ly\alpha \ FWHM}-\rm{Ly\alpha \ EW}$ \citep{kerutt+2022}. 
\end{itemize}

Additionally, simulations have made predictions about correlations between some of the aforementioned quantities, including positive correlations between:
\begin{itemize}
    \item $f_{\rm{esc}}^{\rm{LyC}}-\textrm{Ly$\alpha$ luminosity}$ \citep{kimm+2022, maji+2022} 
    \item and $v_{\rm{sep}}-\textrm{Ly$\alpha$ FWHM}$ \citep{verhamme+2015},
\end{itemize}
and negative correlations between:
\begin{itemize}
    \item $f_{\rm{esc}}^{\rm{LyC}}-v_{\rm{sep}}$ \citep{verhamme+2015, dijkstra+2016, kakiichi&gronke2021, kimm+2022} 
    \item and $v_{\rm{sep}}-\textrm{Ly$\alpha$ EW}$ \citep{verhamme+2015}.
\end{itemize}

To compare our results with previous work, we measured the Pearson correlation coefficient $r$ and type `b' Kendall rank correlation coefficient $\tau$ between all combinations among the Ly$\alpha$ parameters and $f_{\rm{esc}}^{\rm{LyC}}$ for each iteration of the Monte Carlo simulation (Table \ref{tab:param_correlations}). %Because we did not measure the uncertainties in $f_{\rm{esc}}^{\rm{LyC}}$ with a Monte Carlo simulation, 
We incorporated uncertainty in $f_{\rm{esc}}^{\rm{LyC}}$ into the measurement of correlations for each iteration by randomly sampling $f_{\rm{esc}}^{\rm{LyC}}$ according to a Gaussian distribution, where the reported value and uncertainty listed in Table \ref{tab:f_esc} correspond to the mean and standard deviation% of the Gaussian distribution
. We found $f_{\rm{esc}}^{\rm{LyC}}$ strongly correlated with the following Ly$\alpha$ parameters: $v_{\rm{sep}}$, $f_{\rm{min}}/f_{\rm{cont}}$, Ly$\alpha$ EW, and anticorrelated with the blueshifted Ly$\alpha$ peak FWHM (Table \ref{tab:param_correlations}). Between Ly$\alpha$ parameters, we found the following strong correlations:
\begin{itemize}
    \item $\rm{Ly\alpha \ EW}-$$f_{\rm{min}}/f_{\rm{cont}}$, 
    \item $f_{\rm{cen}}-f_{\rm{min}}/f_{\rm{cont}}$, 
    \item $f_{\rm{cen}}-\rm{Ly\alpha \ EW}$, 
    \item $\rm{Ly\alpha \ luminosity}-$$f_{\rm{min}}/f_{\rm{cont}}$,
    \item $\rm{Ly\alpha \ luminosity}-\rm{Ly\alpha \ EW}$,
    \item and $\rm{Ly\alpha \ luminosity}$$-f_{\rm{cen}}$ (Table \ref{tab:param_correlations}).
\end{itemize}
We also found an anticorrelation between $f_{\rm{cen}}$ and $v_{\rm{sep}}$. Apart from the correlation between $f_{\rm{esc}}^{\rm{LyC}}$ and $v_{\rm{sep}}$, these correlations generally align %in line 
with the previous works mentioned above. %We report no strong anticorrelations between any of the parameters, though there may be a moderate anticorrelation between $f_{\rm{cen}}$ and $v_{\rm{sep}}$ ($r=-0.5_{-0.1}^{+0.1}$, $\tau=-0.3_{-0.1}^{+0.1}$)
\begin{figure*}[ht!]
    \centering
    \includegraphics[width=\textwidth]{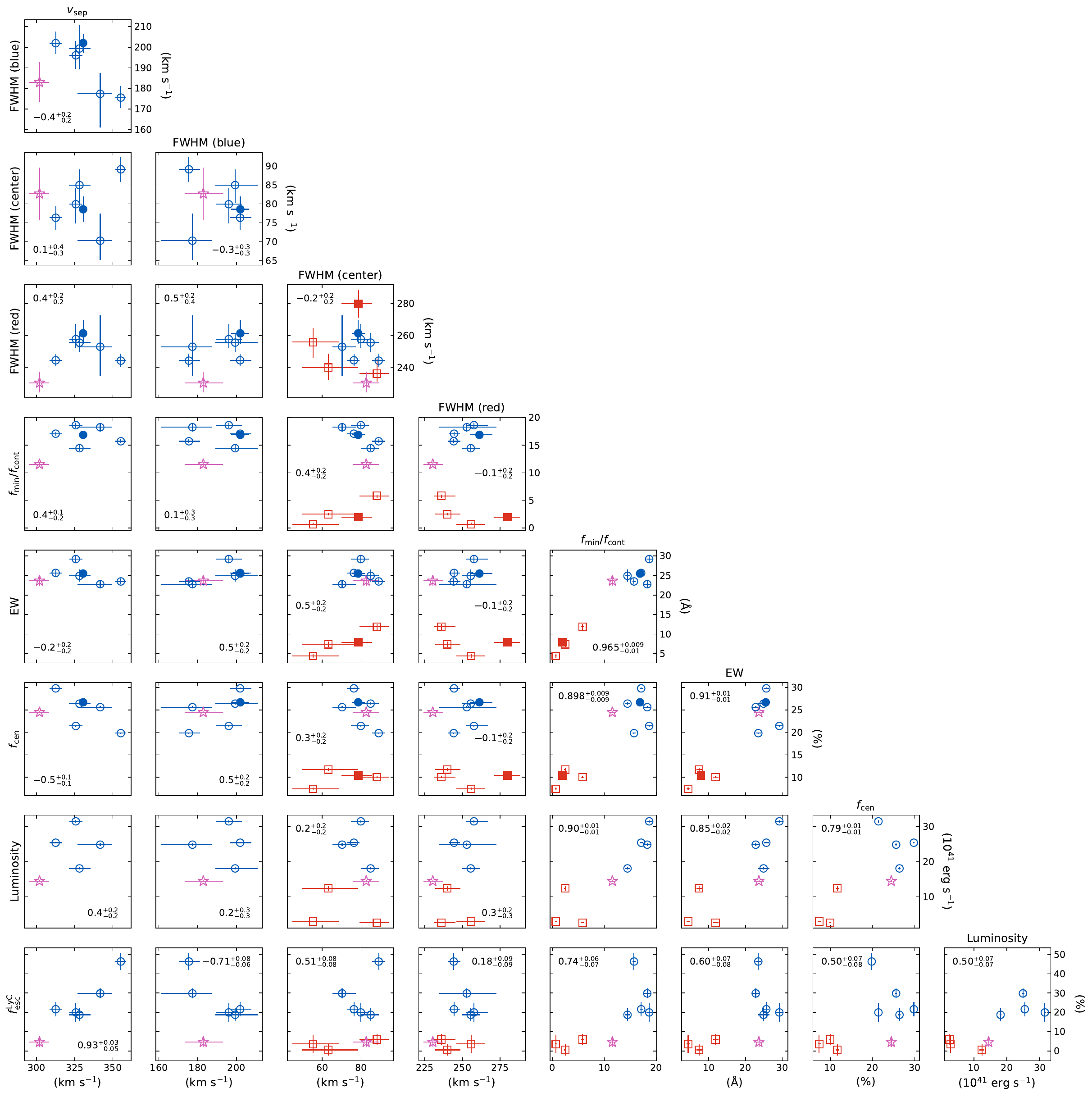}
    \caption{Corner plot of the Ly$\alpha$ measurements, as well as $f_{\text{esc}}^{\text{LyC}}$. Red squares are non-LyC-leaking apertures, blue circles are LyC-leaking apertures, and the pink star is slit M3. Filled markers are the stacked spectra. The value annotated in each plot indicates the Pearson correlation coefficient between the two quantities. Table \ref{tab:param_correlations} also lists the Pearson correlation coefficients and type `b' Kendall rank correlation coefficients.}
    \label{fig:corner}
\end{figure*}

Surprisingly, the Ly$\alpha$ peak separation $v_{\rm{sep}}$ and $f_{\rm{esc}}^{\rm{LyC}}$ show a very significant positive %---not negative---
correlation ($r=0.93_{-0.05}^{+0.03}$), %. This is 
contrary to the %overwhelming 
literature consensus %in the literature 
of a negative correlation between the (galaxy-integrated) $v_{\rm{sep}}$ and $f_{\rm{esc}}^{\rm{LyC}}$. This unexpected result may be due to several factors. %We suggest a few items to consider in response to this unexpected result. 
First, the peculiar shapes of some of the blueshifted Ly$\alpha$ peaks (namely slits M0, M2, and M9 in Figure \ref{fig:lya_fits}, which appear slightly asymmetric relative to the fitted Ly$\alpha$ peak's center) near their center may obfuscate the determination of an accurate Ly$\alpha$ peak separation since we have assumed their skewed Gaussianity (see \S\,\ref{ssec:results_lya} for more discussion). Furthermore, the dynamic range of $v_{\rm{sep}}$ measured from the spectra is minimal and slit M3 impacts the correlation as a clear outlier to the other data points (Figure \ref{fig:corner}). The resolutions of the spectra ($R\gtrsim5000$) are sufficiently high that they may be beginning to resolve departures from the skewed Gaussianity assumption, and thus also the simple shell model geometry and kinematics typically invoked to explain double-peaked LAEs. And because the spectra target small source plane areas, they should be more sensitive to any local departures from an idealized shell model. Also, % consider that 
the established $v_{\rm{sep}}-f_{\rm{esc}}^{\rm{LyC}}$ anticorrelation paradigm has mostly been built from double-peaked LAEs that suggest %minimal directly-escaping Ly$\alpha$ and thus 
a much different mode for LyC escape---where LyC photons may escape because the optical depth of \ion{H}{1} is sufficiently low that many LyC photons do not interact with the \ion{H}{1}. This is not the dominant mode of LyC escape in the Sunburst Arc, where LyC photons primarily escape through an extremely underdense, thin channel \citep{rivera-thorsen+2017, kim+2023}%, so it may be hasty to expect the same trends built from other LyC escape modes to hold. 
. %Confirming the certainty of this tentative correlation between $f_{\rm{esc}}^{\rm{LyC}}$ and $v_{\rm{sep}}$ and precisely explaining it are questions we defer to future work.

%Notably, 
We found some moderate correlations and a strong anti-correlation among the measurement pairs including the FWHM of a Ly$\alpha$ peak, but no strong correlations (Table \ref{tab:param_correlations}). %Considering slit M5's outlier central Ly$\alpha$ peak FWHM, most %of these 
%moderate correlations are likely noncorrelations (Figure \ref{fig:corner}, Table \ref{tab:param_correlations}) %Remarkably, 
We measured very similar FWHMs for all the redshifted %Ly$\alpha$ peaks, and for all the 
and central Ly$\alpha$ peaks (excluding slits M4 and M5 in the latter, which had central Ly$\alpha$ peak FWHMs not much larger than the instrumental dispersion FWHM of $\sim55$ km s$^{-1}$) (Table \ref{tab:lya_params}). %This is interesting since 
The width of a Ly$\alpha$ peak should be sensitive to the number of scattering events the signal experiences before escaping the galaxy, %so there are theoretical grounds to expect 
but we found no clear correlations between the Ly$\alpha$ peak FWHM %to correlate with 
and other proxies for the \ion{H}{1} scattering environment%, but we observe no such clear correlations. 
. However, the FWHM measurements suffer from fewer data points (indicating where a certain Ly$\alpha$ peak is not present in a profile). %The poorly constrained center FWHM of slit M5 is an outlier compared to other center FWHMs, which causes moderate correlations for many comparisons involving the center FWHM, even though the other center FWHMs are remarkably similar (Table \ref{tab:lya_params}). 
Additionally, there is little dynamic range in any of the FWHMs, which fundamentally limits the data's sensitivity to any correlations between FWHMs and other quantities. %could indicate that there are no obvious correlations because the sample probes a small parameter space. %Ultimately, answering specifically how the FWHMs of each Ly$\alpha$ peak type are so similar across the spectra is beyond the scope of this work.

\begin{figure*}
    \centering
    \includegraphics[width=\textwidth]{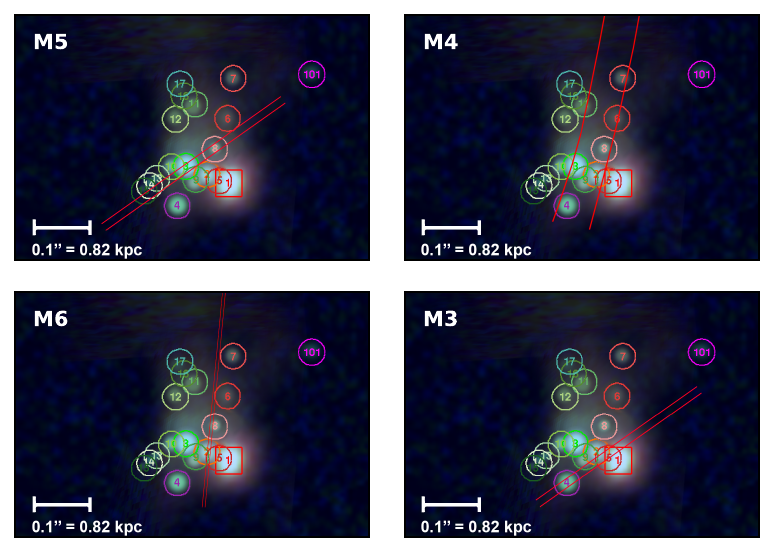}
    \caption{A portion of the approximate source plane reconstruction of the Sunburst Arc from \citet{sharon+2022} with ray-traced non-LyC-leaking apertures overlaid in red and labeled in the top left of each panel. The numbered circles and square correspond to the unique source plane regions identified in \citet{sharon+2022}. Region 1 boxed in red is the LyC leaker.}
    \label{fig:source_plane}
\end{figure*}

The strength of all the measured correlations are subject to caveats. In total, the spectra only target $\lesssim10$ distinct regions in the galaxy's source plane (Figure \ref{fig:source_plane}). Of those regions, the LyC-leaking apertures capture one region, and the non-LyC-leaking apertures capture the remaining regions. But based upon the source plane reconstruction (Figure \ref{fig:source_plane}), the non-LyC-leaking apertures (perhaps excluding slit M6 based upon its geometry) could each include significant contributions from multiple regions of the galaxy. This complicates the interpretation of their Ly$\alpha$ properties since the light in the apertures are the sum of multiple galaxy environments. Also, the LyC-leaking apertures dominate the sample size of many of the parameters and tend to cluster together since they target the same source plane object. This means slit M3, often an outlier to the LyC-leaking region's sample despite also showing a triple-peaked Ly$\alpha$ profile (Figure \ref{fig:corner}), can greatly impact the apparent relation between two measurements.

Additionally, with enough measurements compared against each other, spurious correlations are virtually certain (dubbed data dredging, data snooping, or $p$-hacking). We made 36 comparisons between 9 measurements, with few data points ($5-11$ for any given comparison), so we cannot entirely discount a false detection of a correlation at the measured uncertainties.

\subsection{\texorpdfstring{$f_{\text{esc}}^{\text{LyC}}$}{LyC escape fraction}} \label{ssec:results_fesc}

We found higher $f_{\text{esc}}^{\text{LyC}}$ in the MagE apertures targeting images \citet{rivera-thorsen+2019} identified as LyC leakers ($\gtrsim 20\%$) than those targeting non-LyC-leaking images ($\lesssim 5\%$) (Table \ref{tab:f_esc}). Despite the larger aperture size and PSF of the MagE observations compared to the data and methodology of \citet{rivera-thorsen+2019}, the $f_{\text{esc}}^{\text{LyC}}$ of the LyC-leaking apertures are comparable to or only slightly lower than what \citet{rivera-thorsen+2019} reported. For comparison, \citet{rivera-thorsen+2019} reported a median absolute, IGM absorption-corrected LyC escape fraction from their apertures on the LyC-leaking region of $\approx32\%$. Table \ref{tab:f_esc} summarizes the measured HST photometry and $f_{\rm esc}^{\rm LyC}$.

As expected, the non-LyC-leaking apertures show no obvious LyC emission in the F275W image (Figure \ref{fig:lya_and_lyc_maps}) and minimal $f_{\text{esc}}^{\text{LyC}}$ in our measurements, which use deeper rest-LyC imaging than \citet{rivera-thorsen+2019} (5413 s versus 86902 s). The F275W image of the LyC-leaking images reveals that the LyC emission is extremely compact (Figure \ref{fig:lya_and_lyc_maps}), and at the resolution of HST, the non-LyC-leaking slit apertures do not cover any images of the LyC leaker. Due to the convolution of the data with the ground-based seeing conditions (\S\,\ref{ssec:methods_fesc}), some non-LyC-leaking images may have smaller $f_{\text{esc}}^{\text{LyC}}$ than suggested by this work's aperture-based measurements because the simulation can cause LyC flux from nearby LyC-leaking images to enter the aperture. Since so little LyC flux should already be in a non-LyC-leaking aperture before the convolution, even a small amount could cause a large proportional change, but this likely only affects slit M3, and to a lesser extent slit M4, due to their proximity to images of the LyC-leaking region (Figure \ref{fig:arcs_mage}).

The LyC-leaking apertures show similar or slightly lower $f_{\text{esc}}^{\text{LyC}}$ than %presented for the corresponding images 
the corresponding median absolute, IGM absorption-corrected LyC escape fraction measurement in \citet{rivera-thorsen+2019}. We suggest two factors that may cause this reduction. First, the $f_{\text{esc}}^{\text{LyC}}$ calculation is inversely proportional to the F814W flux. Since our aperture (in effect of order $\sim0.85\arcsec$ (slit width) $\times$ $1.7\arcsec$ (radial arc width)) is much larger than what \citet{rivera-thorsen+2019} used ($\sim0.12\arcsec\times0.12\arcsec$), it is more sensitive to the F814W emission, which is more extended than the compact LyC emission in F275W. This means the F814W flux can effectively dilute the $f_{\text{esc}}^{\text{LyC}}$. Second, the convolution of the data with the ground-based seeing conditions (\S\,\ref{ssec:methods_fesc}) % discussed in the previous paragraph for nonleaker apertures 
introduces an artificial aperture loss: some LyC flux leaves an aperture due to the PSF of the ground-based seeing conditions, decreasing $f_{\text{esc}}^{\text{LyC}}$. This affects the LyC-leaking apertures more significantly due to the comparably-sized slit width and PSF. %since the slit width (0.85\arcsec or 1\arcsec) is comparable to the PSF of the seeing conditions. 
A possible exception is slit M2 (and slit M8 less so), where the slit placement's immediate proximity to the LyC-leaking image 1.9 (Figure \ref{fig:arcs_mage}) suggests nontrivial LyC flux from image 1.9 enters its aperture, even though slit M2 does not directly cover it. This is reflected in slit M2's $f_{\text{esc}}^{\text{LyC}}$, second in strength only to slit M0. A similar process may happen in slit M0 as LyC flux from image 1.6 of the LyC-leaking region enters the aperture. Slit M0 may have an exceptionally high LyC escape fraction because it covers 2 images of the LyC-leaking region---slit M0's LyC escape fraction is also roughly double the LyC escape fractions of the other LyC-leaking apertures.%, but this is marginal compared to the significant contribution from a LyC-bright foreground galaxy (see the diffuse emission SE of image 5 in Figure \ref{fig:arcs_mage}).

\input{lya_measurements_statistical_correlations_table.tex}

\subsection{Ly\texorpdfstring{$\alpha$}{man-alpha} properties} \label{ssec:results_lya}

Our Ly$\alpha$ fitting (\S\,\ref{sssec:methods_lya_fwhm_vsep}) well replicates the overall structure of each velocity profile (Figure \ref{fig:lya_fits}, \ref{fig:lyastack}), but there are some residuals. Deviations from Gaussian or skewed Gaussian behavior are noticeable at the high spectral resolution and high signal-to-noise ratio (SNR) of the spectra, and likely reflect the fact that the outflowing \ion{H}{1} gas has a much more complex geometry than the assumed isotropic shell used to justify the ad-hoc Gaussian and skewed Gaussian model. Key differences include the following:

\begin{itemize}

    \item There is some excess flux in the red tails of the profiles not captured by the fits (Figures \ref{fig:lya_fits}, \ref{fig:lyastack}). These features might reflect velocity intervals in the outflowing gas that have slightly higher column densities. 
    
    \item The skewed Gaussian model fits to the Ly$\alpha$ peaks are not able to fully reproduce the shapes of the spectra. The strongest residual appears in slit M8's redshifted Ly$\alpha$ peak fit (Figure \ref{fig:lya_fits}), in which the data exhibit a sharper decline than the fit can produce. The impact of this effect on the line profile measurements is primarily an additional, small systematic error in the fitted FWHMs, biasing them slightly high.
    
  %  fits assume  may overestimate the redshifted Ly$\alpha$ peak FWHMs, as the observed redshifted Ly$\alpha$ peak in many spectra, compared to the fitted model, falls off faster initially, meets the fitted model by $\sim500$ km s$^{-1}$, and then falls off slower in the redward tail. This is most obvious in slit M8 (Figure \ref{fig:lya_fits}), though we estimate this doesn't cause a difference between the true and measured redshifted Ly$\alpha$ peak FWHM of more than $\sim1$ velocity bin, or $\sim20$ km s$^{-1}$. Something similar may occur to the blueshifted Ly$\alpha$ peak FWHMs, but in reverse, since it seems the blueward wing of the blueshifted Ly$\alpha$ peaks falls off \textbf{faster} than the fits at $\sim-220$ km s$^{-1}$.

    \item The centers of the redshifted Ly$\alpha$ peaks are often broader than the fits. This is especially clear in slits M6, M7, M8, and M9 (Figure \ref{fig:lya_fits}). The minor emission peaks reported by \citet{solhaug+2024} at $\approx130$ km s$^{-1}$ in high-resolution MIKE spectra (Figure 6 therein) may cause this, as the comparatively much lower resolution MagE spectra do not fully resolve these features.
    
    \item An absorber centered at $\sim400$ km s$^{-1}$ attenuates the high-velocity tails of the blueshifted Ly$\alpha$ peaks, as also reported in \citet{solhaug+2024} with much higher resolution spectra. Additionally, the peak structure in several blueshifted Ly$\alpha$ peaks near their center seems more complex than a modest number of Gaussian or skew-Gaussian components can reproduce (particularly slits M0, M2, and M9 in Figure \ref{fig:lya_fits}). %Whether either of these differences are due to intrinsic characteristics of the underlying \ion{H}{1} gas, or connected to absorption effects suggested by the trough at $v\sim-400$ km s$^{-1}$ the spectra also show, is unclear.
    
\end{itemize}

We also note that the structure of the redshifted Ly$\alpha$ peak is remarkably consistent across all of the spectra, indicating that it is likely a ``global'' feature. The remainder of this section discusses the measurements of the individual Ly$\alpha$ parameters. Table \ref{tab:lya_params} summarizes the results.

\subsubsection{Equivalent width}
\label{sssec:results_lya_ew}

The Ly$\alpha$ EWs of the LyC-leaking apertures ($\sim22-30$ {\AA}) are higher than those of the non-LyC-leaking apertures ($\sim4-12$ {\AA}) by a factor of $\gtrsim2$ (excepting slit M3; $24_{-1}^{+1}$ {\AA}) (Table \ref{tab:lya_params}). Since Ly$\alpha$ photons primarily originate from \ion{H}{2} regions, this strong difference is consistent with the much younger age of the LyC-leaking population ($3.3\pm0.5$ Myr according to \citet{mainali+2022}). %Because of its younger age, we expect it should have stronger ionizing flux to sustain the necessary population of \ion{H}{2} ions to recombine and emit Ly$\alpha$ photons.

\subsubsection{Luminosity}

We found $\text{log}_{10}(L_{\text{Ly}\alpha}(\text{erg s}^{-1}))\sim41.4-42.2$ in non-LyC-leaking apertures and $\sim42.2-42.5$ in LyC-leaking apertures (Table \ref{tab:lya_params}). The relationship between Ly$\alpha$ and LyC flux is not always clear. In principle, they are at odds, since LyC photons absorbed by \ion{H}{1} increase the supply of \ion{H}{2} ions that can recombine and emit Ly$\alpha$ photons. But a large $N_{\text{HI}}$ or dust content can effectively suppress both. The extinctions derived by \citet{mainali+2022} suggest the galaxy's dust content does not vary significantly, indicating that older stellar populations in the non-LyC-leaking regions are the dominant effect causing the fainter Ly$\alpha$ luminosities. %in the nonleaking regions are likely their much older (relatively) stellar populations.

\subsubsection{\texorpdfstring{$f_{cen}$}{Central escape fraction}}

We found much higher $f_{\rm{cen}}$ in the LyC-leaking apertures ($\sim 20-30$\%) than non-LyC-leaking ones ($\sim7-10$\%), except for slit M3 ($24.5_{-0.2}^{+0.2}$\%) (Table \ref{tab:lya_params}). \citet{naidu/matthee+22} previously measured $f_{\rm{cen}}\approx37$\% for the LyC leaker in the Sunburst Arc with archival X-SHOOTER data, though the area used to extract the spectrum in that work is unclear. \citet{solhaug+2024} also measured comparatively elevated $f_{\rm{cen}}$ with high-resolution MIKE spectra, finding $f_{\rm{cen}}\approx42$\% for the LyC-leaking region and $f_{\rm{cen}}\approx37$\% for approximately the same pointing as slit M3, which we judged to likely be a consequence of the much higher spectral resolution of their data ($R\approx29,000$). %Our measurements for the LyC-leaking apertures are consistent with other high-$z$, strong LyC leakers \citep{yuan+2021, naidu/matthee+22}.

%and LyC, which are typically in conflict, since extinction of LyC flux by \ion{H}{1} creates more \ion{H}{2} ions which may recombine and emit Ly$\alpha$ photons. The same dilemma appears in the positive correlation between Ly$\alpha$ EW and $f_{\text{esc}}^{\text{LyC}}$ (Figure \ref{fig:corner}). A younger population brighter in LyC wavelengths sustaining a Ly$\alpha$-bright \ion{H}{2} region could resolve this, especially as viewed through a perforating channel into the \ion{H}{2} region. The higher \ion{H}{1} column density to the nonleakers likely compounds this as the thicker \ion{H}{1} rescatters and weakens any Ly$\alpha$ profile emerging from the nonleaker populations.

\subsubsection{Peak widths and separation}

Interestingly, the central and redshifted Ly$\alpha$ peak FWHMs are remarkably similar across most spectra (Table \ref{tab:lya_params}). The central Ly$\alpha$ peak FWHMs ($\sim70-90$ km s$^{-1}$) are also consistent with the results of \citet{solhaug+2024}, who measured central Ly$\alpha$ peak FWHMs of $\sim71-95$ km s$^{-1}$ with the high-resolution ($R\approx29,000$) Magellan/MIKE spectrograph for apertures approximately corresponding to slits M2, M3, and M0. %Only slits M0, M4 and M5 show markedly different central Ly$\alpha$ peak FWHMs %, though at the measured uncertainty it is unclear if even slit M4 is significantly dissimilar compared to the typical central Ly$\alpha$ peak FWHM of $\sim90-100$ km s$^{-1}$.
The redshifted Ly$\alpha$ peak FWHMs are $\sim230-250$ km s$^{-1}$, without significant difference between the LyC-leaking and non-LyC-leaking apertures. Some non-LyC-leaking spectra (specifically slits M3, M4, and M6) might have slightly narrower redshifted Ly$\alpha$ peaks than the LyC-leaking spectra, but the uncertainties make this unclear. If the difference is real, it conflicts with the expectation that the thicker \ion{H}{1} column densities suspected to exist in the non-LyC-leaking regions should broaden emerging Ly$\alpha$ peaks and attenuate LyC flux. However, because the spectra (1) probe small distances in the source plane (much smaller than the galaxy-wide scales of most Ly$\alpha$ observations) and (2) show highly unique Ly$\alpha$ profiles indicative of much different \ion{H}{1} geometry and kinematics than often observed, the same intuitions may not hold in this exotic object. The notable exception is the stacked spectrum of the non-LyC-leaking apertures, which has the broadest redshifted Ly$\alpha$ peak FWHM ($280_{-9}^{+9}$ km s$^{-1}$). This may be connected to possible degeneracies in the model fitting, since the central Ly$\alpha$ peak in the stacked spectrum of the non-LyC-leaking apertures is much fainter than the redshifted Ly$\alpha$ peak. %Also, the redshifted Ly$\alpha$ peak FWHM in the nonleaker apertures may strictly decrease as the relative central Ly$\alpha$ peak strength increases (excepting the stacked nonleaker's Ly$\alpha$ profile), though the measured uncertainties do not make this unambiguous (Table \ref{tab:lya_params}).

The blueshifted Ly$\alpha$ peak FWHMs are not so unusually behaved, but we may detect true differences between them (e.g., slit M0 versus slit M7) due to the larger scatter (Table \ref{tab:lya_params}). If genuine, it is unclear if these differences could be attributable to differential magnification of the images, the viewing angle into the LyC-leaking region, or other effects.

We measured Ly$\alpha$ peak separations $\sim300-350$ km s$^{-1}$ in the LyC-leaking apertures and the non-LyC-leaking aperture slit M3 with a triple-peaked Ly$\alpha$ profile (Table \ref{tab:lya_params}). Low-redshift calibrations compiled by \citet{izotov+2022} suggest $f_{\rm{esc}}^{\rm{LyC}}\lesssim10\%$ based upon these peak separations, much lower than the $f_{\rm{esc}}^{\rm{LyC}}\gtrsim20\%$ we measured for the LyC-leaking apertures (Table \ref{tab:f_esc}). The exception is slit M3, which, despite its narrow peak separation ($302_{-7}^{+6}$ km s$^{-1}$), has a small LyC escape fraction ($f_{\rm{esc}}^{\rm{LyC}}=5\pm3\%$). As noted in \S\,\ref{ssec:results_fesc}, this may be an overestimate due to the simulated seeing effects (\S\,\ref{ssec:methods_fesc}), as an image of the LyC-leaking region is near ($<1\arcsec$) the slit M3 aperture (Figure \ref{fig:arcs_mage}).

\subsubsection{\texorpdfstring{$f_{\rm{min}}/f_{\rm{cont}}$}{Trough-to-continuum ratio}}
\label{sssec:results_lya_fmin_fcont}

We measured $f_{\rm{min}}/f_{\rm{cont}}\sim14-19$ in the LyC-leaking spectra, similar to Green Pea galaxies (GPs) with the narrowest Ly$\alpha$ peak separations \citep{jaskot+2019}. In the non-LyC-leaking spectra, excluding slit M3 ($f_{\rm min}/f_{\rm cont}=11.5_{-0.5}^{+0.4}$), we found $f_{\rm{min}}/f_{\rm{cont}}\sim1-6$, comparable to many GPs with wider peak separations \citep{jaskot+2019}. However, the driving cause of the $f_{\rm{min}}/f_{\rm{cont}}$ measured here is a directly escaping, separate Ly$\alpha$ peak. In most other works %---at least up to their resolving power to constrain the presence of a central Ly$\alpha$ peak---
the measured $f_{\rm{min}}$ is consistent with two superimposed Ly$\alpha$ peaks expected from an isotropically expanding \ion{H}{1} shell. This geometry is not representative of the Sunburst Arc since the central Ly$\alpha$ peak suggests a highly anisotropic ISM.

\section{Discussion} \label{sec:disc}

%\subsection{Ly\texorpdfstring{$\alpha$}{man-alpha} fitting and parameters}

%\subsection{Ly\texorpdfstring{$\alpha$}{man-alpha} spatial and spectral structure}
%\label{ssec:disc_lya}

A key piece in connecting Ly$\alpha$ and LyC escape in the Sunburst Arc is to explain the variety of Ly$\alpha$ profiles, especially the non-LyC-leaking apertures that show a triple-peaked Ly$\alpha$ profile or a central and redshifted Ly$\alpha$ peak. The essential picture explaining the triple-peaked Ly$\alpha$ profile observed in the images of the LyC leaker is well understood. \citet{rivera-thorsen+2017} posited an ionized channel in a surrounding \ion{H}{1} shell as the most likely mechanism to observe a triple peak structure. The much larger $f_{\text{cen}}$, $f_{\text{esc}}^{\text{LyC}}$, and Ly$\alpha$ escape fraction \citep{kim+2023}, and exceptionally blue UV slope \citep{kim+2023} of the LyC-leaking region supports the existence of such a channel oriented along the sightline to the LyC-leaking region. The young age of the LyC leaker suggested by stellar population fitting, in tandem with its strong UV stellar feedback features (e.g., the \ion{N}{5} 1238, 1242 \AA, \ion{C}{4} 1548, 1550 $\text{\AA}$ doublets with P-Cygni wind profiles shown in \citealt{mainali+2022}) and compact size ($r_{\text{eff}}\lesssim32$ pc \citep{sharon+2022}), indicate an obvious culprit with the strong LyC flux and outflows suited to puncture the surrounding \ion{H}{1} medium and leak LyC photons.

What remains unclear are the mechanisms leading to the observed non-LyC-leaking Ly$\alpha$ profiles, since multiple show a central Ly$\alpha$ peak (slits M4, M6, and M3), and in the case of slit M3, a triple-peaked Ly$\alpha$ profile (Figure \ref{fig:lya_fits}). In the channel escape hypothesis, these are both associated with the presence of a LyC source, which do not appear in these apertures (Figure \ref{fig:lya_and_lyc_maps}). 

We have considered if differential magnification effects due to gravitational lensing could explain the observed non-LyC-leaking Ly$\alpha$ profiles. This scenario would require a high magnification gradient, which often occurs close to the critical curve \citep[i.e., Figures 5, 6 in][]{sharon+2022}, and can increase or decrease the weighting of the different physical regions within a lensed galaxy in the spectra that we observe. However, a simple analysis using the lens model \citep{sharon+2022} rejects this as a plausible explanation for the spatially variable Ly$\alpha$ and LyC properties of the Sunburst Arc. In the remainder of this section, we first highlight the peculiar nature of slit M3, which contains a peculiar source and a triple-peaked Ly$\alpha$ profile, but no significant escaping LyC flux. We then discuss possible explanations for all of the observed Ly$\alpha$ profiles in the MagE spectra. 

\subsection{A triple-peaked Ly\texorpdfstring{$\alpha$}{man-alpha} profile without LyC escape}
\label{sssec:disc_lya_m3}

Slit M3 stands out as the only non-LyC-leaking aperture with a triple-peaked Ly$\alpha$ profile, including a blueshifted Ly$\alpha$ peak with a slightly smaller blueshifted velocity than the blueshifted Ly$\alpha$ peak in the LyC-leaking apertures. The brightest continuum source in slit M3 (image 4.8 in Figure \ref{fig:arcs_mage}) has been dubbed \texttt{Tr} \citep{vanzella+2020}, Godzilla \citep{diego+2022,choe+2024,pascale+2024}, and ``the discrepant clump" \citep{sharon+2022}. The literature favors this object as stellar in nature, and we will refer to it as image 4.8, following \citet{sharon+2022}. Here we discuss the Ly$\alpha$ and LyC escape properties of this object. 

A natural starting assumption is to attribute the Ly$\alpha$ emission in slit M3 to the brightest continuum source captured by the slit, which is image 4.8. However, the Ly$\alpha$ narrowband maps from the WFC3/F410M image reveal that image 4.8 itself is mostly likely a net Ly$\alpha$ absorber, with the other diffuse arc flux that falls into slit M3 producing the observed Ly$\alpha$ emission. Inspecting slit M3's position in the source plane reconstruction indicates that the emission captured by slit M3 comes from two physically distinct parts of the galaxy: region 4, and diffuse emission from or between regions 1, 2, and 5. This is notable because this diffuse emission in slit M3 comes from a region that is closer to the LyC-leaking region (but without containing the LyC-leaking region itself) than any other of the MagE slit apertures.

We observe LyC flux in slit M3's aperture statistically consistent with zero flux (at $2\sigma$ confidence; Table \ref{tab:f_esc}). However, the young age ($\sim4$ Myr), low stellar extinction of $\text{E}(\text{B}-\text{V})=0.018$ \citep[the lowest measured in the MagE data;][]{mainali+2022}, and narrow Ly$\alpha$ peak separation suggest a low \ion{H}{1} column density and a favorable stellar population and line of sight for LyC escape. However, it is difficult to tie all of these characteristics to the same physical region, since slit M3 covers emission regions separated by $\sim$1 kpc in the source plane (Figure \ref{fig:source_plane}). Here we review how the different physical natures offered for image 4.8 in the literature connect to the Ly$\alpha$ and LyC observations presented in this paper.

\begin{itemize}

\item \textbf{A supernova.} \citet{vanzella+2020} first discussed the main image's identity, positing it to be a transient stellar object, evidenced by the lack of counterimages, and the lack of any other object with the same unique spectroscopic signatures, namely Bowen fluorescence \citep{bowen1934}. Based on their predicted magnification ($20<\mu<100$) and corresponding absolute magnitude ($-20.3<M_{2000}<-18.6$) of the main image, they favored a supernova and disfavored a luminous blue variable (LBV) star as too faint. However, detailed lens modeling by \citet{diego+2022} and \citet{sharon+2022} jointly corroborated that (1) the magnification of the main image must be extremely high ($\mu>600$), (2) the source plane persistence of the main image is exceptionally long for a supernova ($>1$ year), and (3) the predicted time delays for this source between different images ($<1$ year) are significantly shorter than the elapsed time of observations. The lack of appearance of any new, comparably bright images of ``Godzilla'' over $>7$ years strongly disfavors the supernova hypothesis. The lack of Ly$\alpha$ and LyC emission from image 4.8 is consistent with a SN.

\item \textbf{A luminous blue variable star.} In contrast, \citet{diego+2022} suggested the largest knot in slit M3 could be a luminous blue variable (LBV) star in outburst. This could explain the observed Ly$\alpha$ signature and faint LyC detection, as well as rest-UV spectral features \citep{vanzella+2020}. Quiescent LBVs are often B-type stars with temperatures ranging between $\sim10000-25000$ K, but during their outbursts, they cool to $\sim8500$ K and shed their outer layers as large mass outflows. Assuming the star's quiescent temperature is among the hotter LBVs, the star may produce appreciable LyC flux. The radiative and mechanical energy from the LBV during an active phase could clear an underdense channel in the surrounding \ion{H}{1}, but this picture is in tension with \citet{choe+2024}, who concluded from JWST/NIRSpec IFU observations that the main image is a \textit{quiescent} LBV, not an outbursting LBV.  Furthermore, the dense gas required to pump Bowen fluorescence would naturally suppress escaping Ly$\alpha$ and LyC emission from the LBV, consistent with the observed lack of Ly$\alpha$ and LyC emission from image 4.8

\end{itemize}

As we explore physical hypotheses to explain the observed Ly$\alpha$ profiles below, it is important to remember that any plausible scenario must explain how slit M3 could show both central and blueshifted Ly$\alpha$ peaks but no escaping LyC photons.

\begin{figure*}
    \centering
    \includegraphics[width=\textwidth]{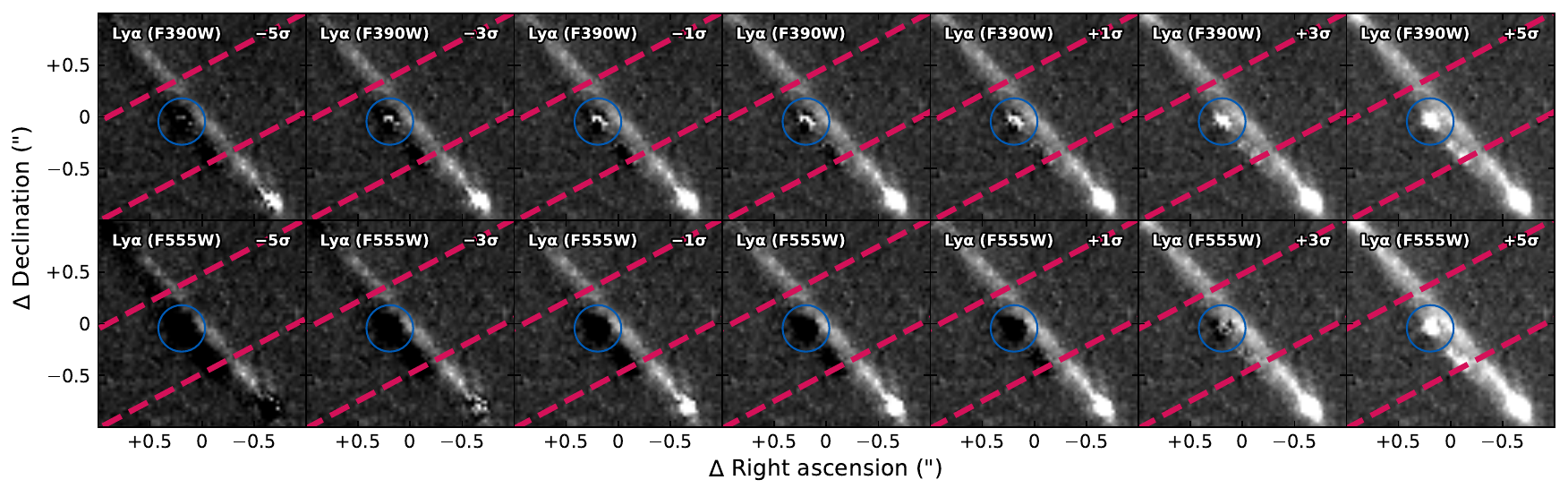}
    \caption{Ly$\alpha$ maps centered on slit M3 created from the schemes discussed in \S\,\ref{ssec:methods_nb_lya}. The top and bottom rows show the maps with the F390W and F555W filters as the off-band filter, respectively. The dashed magenta line is the on-sky aperture of slit M3. There are systematic differences between the two continuum subtraction bands, with the F555W-based subtraction removing marginally more continuum flux than the F390W-based subtraction}.
    \label{fig:m3_lya}
\end{figure*}

\subsection{Seeing effects}
\label{sssec:disc_lya_seeing}

We have considered if ground-based seeing conditions cause the unique kinematic structure of the Ly$\alpha$ profiles of the non-LyC-leaking apertures by diffusing images of the LyC leaker's triple-peaked Ly$\alpha$ profile into non-LyC-leaking apertures. To judge this, we simulated the atmospheric diffusion with the Ly$\alpha$ maps discussed in \S\,\ref{ssec:methods_nb_lya}. For each slit aperture, we convolved each Ly$\alpha$ map with a 2-dimensional Gaussian kernel with width matching the time-weighted average of the combined effect of the seeing conditions (see Table \ref{tab:mage_log} for a time-weighted average of the seeing conditions for each slit) and airmass for each exposure (Table 1). We then calculated the flux inside each slit aperture before and after the convolution and the ratio between those fluxes for each Ly$\alpha$ map. Table \ref{tab:seeing_simulation} contains the results. 

In both the Ly$\alpha$ maps made from the F555W- and F390W-based continuum subtraction, we found that the LyC-leaking apertures lose Ly$\alpha$ flux ($\sim10-40\%$) and the non-LyC-leaking slits mostly gain Ly$\alpha$ flux ($\sim 5-800\%$). The one exception is slit M5, which has a pre- and post-convolution flux that is statistically consistent to within $\sim2\sigma$ in both continuum subtraction schemes.

We now outline how each non-LyC-leaking aperture fits into the seeing effects hypothesis, roughly in order of the apertures nearest to farthest from images of the LyC-leaking region on the sky.

\begin{itemize}
    \item[\textbf{M3}] Slit M3 is roughly the non-LyC-leaking aperture nearest to images of the LyC-leaking region (Figure \ref{fig:lya_and_lyc_maps}) and the only non-LyC-leaking aperture that has a blueshifted Ly$\alpha$ peak. This slit is the most consistent with the seeing effects hypothesis since it shows the hallmark triple-peaked Ly$\alpha$ profile of the LyC leaker and is $<1\arcsec$ away from the nearest image of the LyC-leaking region. Slit M3's Ly$\alpha$ flux either changes significantly or minimally in the simulated seeing convolution, depending on the Ly$\alpha$ map (Table \ref{tab:seeing_simulation}). This ambiguity is because the F555W-based Ly$\alpha$ map predicts strong Ly$\alpha$ absorption near and on the images in slit M3 (Figure \ref{fig:m3_lya}). Additionally, slit M3's blueshifted Ly$\alpha$ peak is distinctly closer to the systemic redshift than the blueshifted Ly$\alpha$ peaks observed from the LyC-leaking region (Figure \ref{fig:lya_fits}, Table \ref{tab:fit_results}), which point to a different physical origin for slit M3's blueshifted Ly$\alpha$ peak than simple seeing effects.

    \item[\textbf{M4}] Slit M4 is farther from the nearest images of the LyC-leaking region than slit M3, but still close enough ($\sim1\arcsec$) that seeing effects could cause some contamination (Figure \ref{fig:lya_and_lyc_maps}). Similar to slit M3, the Ly$\alpha$ flux in the aperture either changes significantly or minimally after the simulated seeing convolution, depending on the Ly$\alpha$ map (Table \ref{tab:seeing_simulation}). Most importantly, slit M4 does not show a blueshifted Ly$\alpha$ peak (Figure \ref{fig:lya_fits}), which we would expect to see if significant Ly$\alpha$ flux from the LyC-leaking region diffused into this slit due to seeing effects. 
    
    \item[\textbf{M6}] Slit M6 is well-separated from the closest images of the LyC-leaking region (Figure \ref{fig:lya_and_lyc_maps}), but still has a strong central Ly$\alpha$ peak, though no blueshifted Ly$\alpha$ peak. At most, the seeing-convolved Ly$\alpha$ images predict a moderately increased Ly$\alpha$ flux due to seeing effects. The absolute, magnification-uncorrected strength of slit M6's central Ly$\alpha$ peak is several times greater than that of slit M4 (Table \ref{tab:fit_results}), even though slit M4 is closer to images of the LyC-leaking region. This fact (and the nonobservation of a blueshifted Ly$\alpha$ peak) strongly disfavors the seeing effects hypothesis since an aperture (slit M6) farther from images known to emit a central Ly$\alpha$ peak has a stronger central Ly$\alpha$ peak than another, closer aperture (slit M4).

    \item[\textbf{M5}] Slit M5 shows no distinct central or blueshifted Ly$\alpha$ peak and is one of the non-LyC-leaking apertures farthest from images of the LyC leaker (Figure \ref{fig:lya_and_lyc_maps}), which is consistent with the seeing effects hypothesis. Like slit M6, the simulated seeing convolution predicts a minimal or moderate increase in Ly$\alpha$ flux in the slit M5 aperture due to the observing conditions (Table \ref{tab:seeing_simulation}).
    
\end{itemize}

\input{seeing_simulation_measurements_table.tex}

Combined, the evidence above suggests seeing effects alone cannot explain all the Ly$\alpha$ profiles of the non-LyC-leaking apertures, but might affect their Ly$\alpha$ profiles in some instances---such as slit M3 due to its proximity to images of the LyC-leaking region. Because both ground-based seeing effects (\S\,\ref{sssec:disc_lya_seeing}) and differential magnification from gravitational lensing do not explain the observed Ly$\alpha$ profiles of the non-LyC-leaking apertures, we must consider explanations that invoke complex mechanisms and morphologies, which we explore below. 

\subsection{Multiple direct escape Ly$\alpha$ channels}
\label{sssec:disc_lya_channels}

Here we consider if additional ionized channels might also explain the central Ly$\alpha$ peaks observed from the non-LyC-leaking apertures. Any channel escape scenario must explain the observation of a central Ly$\alpha$ peak but not a blueshifted Ly$\alpha$ peak in some of the non-LyC-leaking apertures (slits M4 and M6). If the additional ionized channels are similar to the channel observed in the LyC-leaking region (i.e., additional small channels puncturing an isotropically expanding \ion{H}{1} shell), then the gas along the sightlines toward the additional channels must have properties that permit Ly$\alpha$ photons to directly escape, but not LyC photons. 

Such a scenario is difficult to construct because the cross section for Ly$\alpha$ photons to interact with \ion{H}{1} is $\sim10^4$ times higher than the cross section for LyC photon interactions. This means that if there is sufficiently low \ion{H}{1} column density along a sightline to allow Ly$\alpha$ photons to escape with few or no scattering events (``direct escape'' Ly$\alpha$), then LyC photons should also freely escape. Furthermore, a blueshifted Ly$\alpha$ peak is associated with outflowing gas with low \ion{H}{1} column density. The lack of a blueshifted Ly$\alpha$ peak in some of the non-LyC-leaking apertures with a central Ly$\alpha$ peak requires either (1) simultaneously both high column density, outflowing \ion{H}{1} and an extremely low column density \ion{H}{1} channel puncturing it, or (2) a complete absence of outflowing \ion{H}{1}. The former would require a sharp transition between the two gas phases to minimize the \ion{H}{1} thin enough to transmit a blueshifted Ly$\alpha$ peak but thick enough to prevent direct Ly$\alpha$ scape, while the latter would allow LyC photons to efficiently escape (which we do not observe).

We also note that the stellar populations in the non-LyC-leaking regions of the Sunburst Arc have ages of $11.8\pm0.9$ Myr \citep{mainali+2022}, which is notably older and should produce fewer LyC photons \citep{ma+2015, ma+2020, kimm+2017, kim+2019, kakiichi&gronke2021} than the LyC leaking region's age of $3.3\pm0.5$ Myr \citep{mainali+2022}. The rest-UV absorption lines and nebular emission lines in the non-LyC-leaking spectra also indicate weaker stellar winds, bulk outflows \citep{mainali+2022}, and less extreme ionizing radiation fields \citep{kim+2023} than the LyC-leaking region, meaning the non-LyC-leaking regions are less well-suited to create ionized channels. This makes it physically unlikely for the non-LyC-leaking regions to be able to create additional highly ionized channels, especially considering that radiative transfer simulations predict channels to be extremely rare \citep[e.g.,][]{behrens+2014}.

\subsection{A Ly\texorpdfstring{$\alpha$}{man-alpha} mirror}
\label{sssec:lya_mirror}

We now introduce a Ly$\alpha$ `mirror' hypothesis, which posits that dense \ion{H}{1} cores could preferentially scatter Ly$\alpha$ photons from the LyC-leaking region into our sightline far from their origin. As discussed in \S\,\ref{sec:intro}, Ly$\alpha$ photons may wander far from their birthplace before scattering into our sightline. The central Ly$\alpha$ peak observed in some of the non-LyC-leaking apertures might be an extreme case of this: Ly$\alpha$ photons of the LyC-leaking region incident upon dense \ion{H}{1} gas surrounding the non-LyC-leaking regions may preferentially scatter off of the cores' surfaces and cause apparent Ly$\alpha$ emission from regions physically distant from the original Ly$\alpha$ source.

\begin{figure}
    \centering
    \includegraphics[width=\columnwidth]{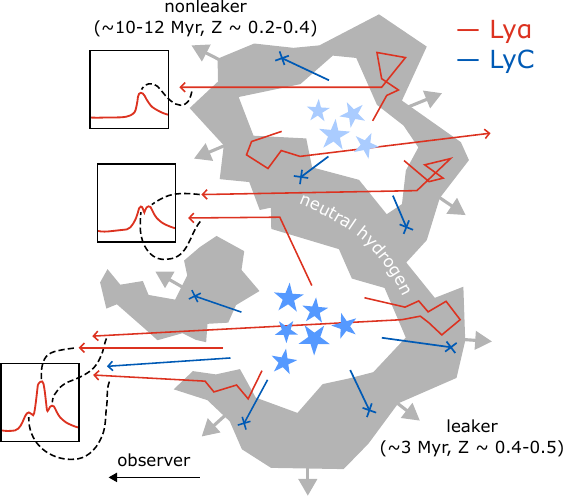}
    \caption{
    Cartoon of the Ly$\alpha$ mirror hypothesis. The LyC-leaking region (bottom) produces a prodigious amount of Ly$\alpha$ (red) and LyC (blue) photons compared to the older non-LyC-leaking regions (top). Some photons from the LyC-leaking region directly escape out of a highly ionized channel oriented toward us. These photons constitute the observed LyC and central Ly$\alpha$ peak. Additional Ly$\alpha$ emission resonantly scatters off the expanding \ion{H}{1} surrounding the LyC-leaking region until it escapes the galaxy along the observer’s sightline, which constitute the redshifted and blueshifted Ly$\alpha$ peaks (bottom box). In the non-LyC-leaking regions, which have much less LyC flux due to their older age, the surrounding \ion{H}{1} (gray) is thicker, preventing any LyC escape and direct Ly$\alpha$ escape. Thus, only a redshifted Ly$\alpha$ peak emerges (first box from the top). However, Ly$\alpha$ photons from the LyC-leaking region preferentially reflect into the observer’s sightline incident upon the thick \ion{H}{1} around the non-LyC-leaking regions (second box from the top), forming a central Ly$\alpha$ peak superimposed upon the redshifted Ly$\alpha$ peak emerging from the non-LyC-leaking regions.}
    \label{fig:mirror_cartoon}
\end{figure}

This preferential scattering is similar to the mechanism that produces the redshifted Ly$\alpha$ peak in an isotropically expanding \ion{H}{1} shell geometry \citep[e.g., Figure 12 in][]{verhamme+2006}. If this process involved only a few scatterings then the central Ly$\alpha$ peak may survive, albeit diminished, which is what we observe in the non-LyC-leaking apertures that are closest to the LyC-leaking region in the source plane (Figure \ref{fig:lya_fits}). This would require either a large region with very low column density \ion{H}{1} and/or very fast (i.e., out of resonance) outflowing gas between the ``mirror'' and our sightline to allow the central Ly$\alpha$ photons to escape. The strong ionizing photon flux of the LyC-leaking region may aid this process by creating a large ionized region inside the galaxy, allowing Ly$\alpha$ photons to traverse large distances far from their birth place. Figure \ref{fig:mirror_cartoon} sketches out this hypothesis.

We now outline how this hypothesis connects to the Ly$\alpha$ profiles of each non-LyC-leaking aperture, in order of the strongest to weakest central Ly$\alpha$ peaks, relative to their associated redshifted Ly$\alpha$ peaks.%, and what makes.

\begin{itemize}
    \item[\textbf{M3}] This aperture covers parts of regions 1, 2, 4, 5, and 9 in the source plane (Figure \ref{fig:source_plane}). Slit M3 is the closest non-LyC-leaking aperture to the LyC-leaking region, meaning some of the area it covers---namely regions 1, 2, and 5---could receive the necessary Ly$\alpha$ flux to scatter into our sightline before significant Ly$\alpha$ photons scatter prematurely, attenuate, or become too diffuse. If a Ly$\alpha$ `mirror' causes slit M3's strong central Ly$\alpha$ peak, there must be very few scatterings and a strong directional bias toward us to preserve a central Ly$\alpha$ peak that is nearly as strong (relative to the spectrum's other Ly$\alpha$ peaks) as in the LyC-leaking apertures. There must also be very low column density outflowing \ion{H}{1} in this region to explain the presence of a blueshifted Ly$\alpha$ peak in slit M3. The lack of LyC photons within this aperture suggests that that there are no bright LyC photon sources in the diffuse emission captured by slit M3, such that the Ly$\alpha$ emission detected from this slit is the product of Ly$\alpha$ photons that diffused away from the LyC-leaking region before scattering into our sightline.

   \item[\textbf{M6}] This aperture covers parts of regions 2, 5, and 8. Slit M6's central Ly$\alpha$ peak is comparable in strength to its redshifted Ly$\alpha$ peak (Figure \ref{fig:lya_fits}) and its projected distance to the LyC-leaking region in the source plane is farther than for slit M3 (Figure \ref{fig:source_plane}). The central Ly$\alpha$ peak of slit M6 could also result from the ``mirror'' effect. Slit M6's larger distance from the LyC-leaking region might explain the weaker central Ly$\alpha$ peak (relative to the redshifted Ly$\alpha$ peak) compared to the LyC-leaking region due to some combination of (1) a smaller covering fraction of the mirror for slit M6 than slit M3, and (2) an increasing number of scattering events for Ly$\alpha$ photons as they travel farther from their original source in the LyC-leaking region.
    
    \item[\textbf{M4}] This aperture targets a lower magnification region of the arc and therefore covers a large portion of the lensed galaxy in the source plane, including regions $2-11$. Though parts of the slit are closer to the LyC-leaking region in the source plane than slit M6 (Figure \ref{fig:source_plane}), its central Ly$\alpha$ peak is much weaker compared to its redshifted Ly$\alpha$ peak than in slit M6, though the magnification-corrected central Ly$\alpha$ flux is still stronger absolutely than that of slit M6 (Figure \ref{fig:lya_fits}, Table \ref{tab:fit_results}). This makes sense given the large footprint of slit M4 in the source plane if the redshifted Ly$\alpha$ peak is a global feature, while the central Ly$\alpha$ peak is not. 
    
    \item[\textbf{M5}] This aperture covers parts of regions 3, $8-10$, and $13-15$. Of all the non-LyC-leak apertures, it has the largest projected distance from the LyC-leaking region in the source plane, and it is the only non-LyC-leaking aperture with no clear central Ly$\alpha$ peak. This implies a minimum upper limit on the extent of the central Ly$\alpha$ peak of $\lesssim800$ pc.

\end{itemize}

If a Ly$\alpha$ mirror does exist, it extends to at least clumps 1, 2, 5, and 9 in Figure \ref{fig:source_plane}, since those clumps define a region of overlap between slits that show a central Ly$\alpha$ peak (M3, M4, and M6). This suggests a ``direct escape'' Ly$\alpha$ region extending as much as $\sim600-800$ pc from the LyC-leaking region.

\begin{figure}
    \centering
    \includegraphics[width=\columnwidth]{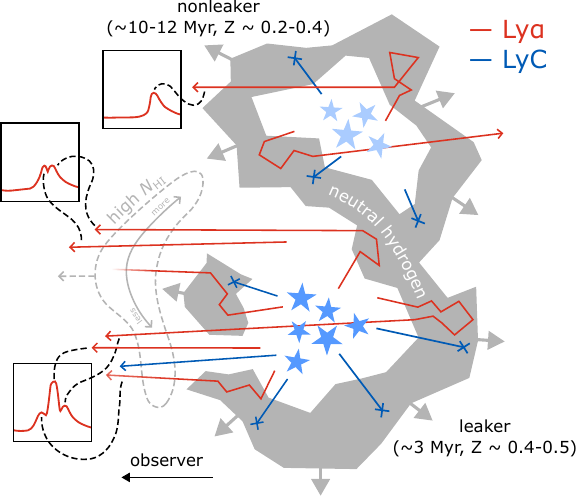}
    \caption{Cartoon sketch of the spatially variable absorption hypothesis. The Ly$\alpha$ and LyC photons directly escaping through an ionized channel produce the LyC and the central Ly$\alpha$ peak observed from the LyC-leaking region. Ly$\alpha$ photons from the LyC-leaking region also resonantly scatter with either side of the surrounding, expanding \ion{H}{1} shell, forming the redshifted and blueshifted Ly$\alpha$ peaks (bottom box). In this picture, the triple-peaked Ly$\alpha$ profile from the LyC-leaking region originates from a large source plane area, and an intervening, outflowing absorber with higher HI column density attenuates the blueshifted Ly$\alpha$ peak and varying amounts of the central Ly$\alpha$ peak across the face of the galaxy.}
    \label{fig:diffused_absorption_cartoon}
\end{figure}

\subsection{Spatially variable Ly$\alpha$ absorption}
\label{sssec:disc_lya_diffuse_absorber}

The final hypothesis we present also relies on a complex \ion{H}{1} morphology in the source plane. We suggest that the triple-peaked Ly$\alpha$ profile of the LyC-leaking region occupies an area in the source plane that is much larger than the clump producing the ionizing radiation (the LyC-leaking region). In this picture, the LyC-leaking region has ionized a huge volume of gas, producing the central Ly$\alpha$ peak over a large extent, as escaping Ly$\alpha$ photons can either form in-situ from the extended ionized gas, or travel large transverse distances in the source plane before experiencing scattering event(s) that redirect them into other, non-LyC-leaking sightlines. 

The spatial variability of the blueshifted and central Ly$\alpha$ peaks' strength is likely due to an outflowing (blueshifted) intervening absorber with \ion{H}{1} column density that varies spatially across the face of the galaxy. For this absorber to screen the central Ly$\alpha$ peak, there must be additional highly ionized sightlines toward the extended ionized gas. For example, the \ion{H}{1} surrounding the extended ionized gas may be patchy and heavily punctured. However, the \ion{H}{1} cannot be significantly perforated, or there would not be sufficient approaching, outflowing \ion{H}{1} to reprocess Ly$\alpha$ photons into the blueshifted Ly$\alpha$ peak that we see in slit M3 and the LyC-leaking apertures. Figure \ref{fig:diffused_absorption_cartoon} depicts the basic principles of this scenario. 

There are several key unknowns involved in this picture. First, what is the size and structure of the highly ionized, extended region around the LyC-leaking region? Second, what is the location and morphology of the neutral \ion{H}{1} gas that absorbs LyC photons from all sightlines except those from a small channel toward the LyC-leaking region? The first question could be addressed in part by future spatially resolved measurements of the physical extent of forbidden (optically thin) nebular emission lines in the source plane. The second question, however, may require very specific \ion{H}{1} morphology and kinematics to efficiently diffuse the directly escaping, central Ly$\alpha$ peak from the LyC-leaking region without completely rescattering it---perhaps not too dissimilar from the \ion{H}{1} morphology and kinematics presumed in the mirror hypothesis of \S\,\ref{sssec:lya_mirror}---if most of the central Ly$\alpha$ peak's photons do not form in-situ from the extended ionized gas.

Significant supporting evidence exists for this hypothesis. For instance, the neutral gas-tracing low ionization absorption analyzed by \citet{mainali+2022} suggests a blueshifted, low-ionization absorber exists along both LyC-leaking and non-LyC-leaking sightlines. That result is also consistent with the absorption trough at $\sim-350$ km s$^{-1}$ (Figure \ref{fig:lya_fits}) that is present in all of the Ly$\alpha$ spectra. Furthermore, this blueshifted absorption trough has noticeably stronger absorption tails in the non-LyC-leaking MagE spectra, suggesting an absorption gradient between the LyC-leaking and non-LyC-leaking regions, with an increasing column density as the slits move farther away from the LyC-leaking region. This absorber might also explain the sharp drop-off on the blue side of the blueshifted Ly$\alpha$ peaks. Below, we explore how each non-LyC-leaking aperture might fit into this picture, in order of increasing projected distance from the LyC-leaking region, based on the source plane reconstruction (Figure \ref{fig:source_plane}) and Ly$\alpha$ profile fits (Figure \ref{fig:lya_fits}).

\begin{itemize}
    \item[\textbf{M3}] Of the non-LyC-leaking apertures, slit M3 is nearest to the LyC-leaking region in the source plane. It also shows the least signs of absorption compared to the LyC-leaking region's Ly$\alpha$ profile, all of which is fully consistent with this physical picture. A very weak absorption process also provides the simplest and clearest explanation for slit M3's blueshifted Ly$\alpha$ peak, which is slightly less blueshifted than the blueshifted Ly$\alpha$ peak in the LyC-leaking spectra. Similar to the Ly$\alpha$ mirror hypothesis above, the lack of LyC photons within this aperture suggests that all of the Ly$\alpha$ photons observed in slit M3 stem from Ly$\alpha$ photons that diffused or originated far from the LyC-leaking region before entering our sightline.

    \item[\textbf{M4}] Slit M4 is the next closest slit to the LyC-leaking region in the source plane, though it also has the largest footprint and captures many non-LyC-leaking regions distant from the LyC-leaking region. Slit M4 shows a suppressed central Ly$\alpha$ peak and no blueshifted Ly$\alpha$ peak. In the gradient absorber picture, this would indicate the absorber's optical depth has grown, which is consistent with the increased distance of slit M4 to the LyC-leaking region in the source plane.

    \item[\textbf{M6}] Slit M6 has a very high magnification ($\mu=147_{-20}^{+5}$, Table \ref{tab:mage_log}), causing the slit to be very narrow in the source plane, isolating region 2. Slit M6 is an interesting contrast to slit M4 (which is one of the least magnified slits). Comparing the total, magnification-corrected flux in the central Ly$\alpha$ peaks of slits M6 and M4, there is significantly more flux in slit M4, but comparing the {\it relative} strengths of the central and redshifted Ly$\alpha$ peaks, it is clear slit M6's central Ly$\alpha$ peak is stronger relative to the redshifted Ly$\alpha$ peak than in slit M4. This makes sense if (1) the redshifted Ly$\alpha$ peak is a global emission feature with approximately constant surface brightness across the entire source plane, and (2) the central Ly$\alpha$ peak emission is localized to a region that extends a few hundred parsecs around the LyC-leaking region. Slit M4 has a much larger footprint that could then contain Ly$\alpha$ profiles from other non-LyC-leaking regions (which slit M6 does not cover) with higher \ion{H}{1} column densities and a more diminished or absent central Ly$\alpha$ peak and stronger absorption trough.

    \item[\textbf{M5}] Slit M5 is the farthest aperture from the LyC-leaking region in the source plane. It has no clear evidence of any residual blueshifted or central Ly$\alpha$ peak. Its Ly$\alpha$ profile could be consistent with a single redshifted Ly$\alpha$ peak. This slit is consistent with the redshifted Ly$\alpha$ peak as a global emission feature that extends across the entire galaxy. If the central Ly$\alpha$ peak's production region extends as far as slit M5 in the source plane, then the intervening absorber along this slit's sightline must have a very high \ion{H}{1} column density to suppress the central Ly$\alpha$ emission. It is also possible that the physical extent of the region emitting the central Ly$\alpha$ peak ends somewhere between slits M6 and M5 in the source plane.
    
\end{itemize}

Combined, we believe that a diffuse triple-peaked Ly$\alpha$ profile combined with blueshifted \ion{H}{1} absorption as a physical scenario has significant merit, especially considering the ubiquitous absorption trough in every Ly$\alpha$ profile.

\input{lya_best_fit_model_parameters_table.tex}

\section{Conclusions} \label{sec:con}

% Summarize main findings
In this work, we analyzed rest-UV, high-resolution ($R\sim5000$) spectroscopy of the Ly$\alpha$ emission line in the Sunburst Arc (\S\,\ref{ssec:obs_mage}), a strongly lensed, LyC-leaking galaxy at $z\approx2.37$. We directly fitted the Ly$\alpha$ velocity profiles to extract key parameters (\S\,\ref{ssec:methods_lya}) such as the peak separation $v_{\rm{sep}}$ (\S\,\ref{sssec:methods_lya_fwhm_vsep}), peak widths (\S\,\ref{sssec:methods_lya_fwhm_vsep}), Ly$\alpha$ equivalent width (\S\,\ref{sssec:methods_luminosity}), central escape fraction $f_{\rm{cen}}$ (\S\,\ref{sssec:f_cen}), and ratio between the minimum flux density between the peaks and local continuum flux density $f_{\rm{min}}/f_{\rm{cont}}$ (\S\,\ref{sssec:methods_lya_fmin_fcont}). Using HST imaging (\S\,\ref{ssec:obs_hst}), we computed the LyC escape fractions associated with the spectroscopic observations (\S\,\ref{ssec:methods_fesc}).

From this, we computed statistical correlations between the Ly$\alpha$ and LyC parameters to better understand their interdependence (\S\,\ref{sec:results}), and offered brief discussion of the measurements of each parameter (\S\,\ref{ssec:results_fesc}, \S\,\ref{sssec:results_lya_ew}--\S\,\ref{sssec:results_lya_fmin_fcont}), and possible shortcomings of the Ly$\alpha$ profile fitting (\S\,\ref{ssec:results_lya}). In short, we found:
\begin{itemize}

    \item[(i)] a strong positive correlation between $f_{\rm{esc}}^{\rm{LyC}}$ and $v_{\rm{sep}}$, $f_{\rm{min}}/f_{\rm{cont}}$, and Ly$\alpha$ EW (Table \ref{tab:param_correlations}),
    
    \item[(ii)] few clear relations between any two parameters involving any of the Ly$\alpha$ peak widths (Table \ref{tab:param_correlations}), and

    \item[(iii)] affirmations of previously suggested relations, especially a strong correlation between the Ly$\alpha$ EW and $f_{\rm{min}}/f_{\rm{cont}}$.
    
\end{itemize}

We also note that a central Ly$\alpha$ peak appears even in spectra of non-LyC-leaking regions from apertures located within $\sim600$ pc of the LyC-leaking region. We briefly discussed the observed spectrum of the highly magnified region known as ``Godzilla'' and its implications for the nature of the object.
Based upon the data and a source plane reconstruction of the slit aperture geometry (Figure \ref{fig:source_plane}), we discussed several possible explanations for the diversity of Ly$\alpha$ profiles observed in the Sunburst Arc (\S\,\ref{sec:disc}), particularly those from the non-LyC-leaking spectra. We first considered if ground-based seeing effects (\S\,\ref{sssec:disc_lya_seeing}) could cause the observed Ly$\alpha$ profiles by diffusing emission from images of the LyC-leaking region into the non-LyC-leaking slit apertures, but concluded that this mechanism cannot explain the data.

We then discussed physical mechanisms to explain the observed Ly$\alpha$ profiles of the non-LyC-leaking apertures (\S\,\ref{sssec:disc_lya_channels}$-$\S\,\ref{sssec:disc_lya_diffuse_absorber}). We concluded that the data are inconsistent with the presence of additional highly ionized channels that permit direct Ly$\alpha$ escape without LyC escape (\S\,\ref{sssec:disc_lya_channels}). We then sketched out two physical scenarios---the Ly$\alpha$ mirror (\S\,\ref{sssec:lya_mirror}) and the spatially variable Ly$\alpha$ absorber (\S\,\ref{sssec:disc_lya_diffuse_absorber})---to explain the presence of a central Ly$\alpha$ peak that originates from a much larger physical region than the escaping LyC photons.
%Because the available spectroscopic data poorly constrains the number or locations of additional ionized channels, we suggest the 
%multiple channels aligned in a single direction is statistically unlikely without a clear mechanism to create and align the channels along a common sightline, (2) the nonleaking stellar populations do not appear to have the strong ionizing flux and energetic outflows that promote channel creation. Such a speculative mechanism is poorly constrained because the spectroscopic MagE data does not clearly constrain how many additional, non-LyC-leaking channels could exist, or where those channels are in the source plane. We concluded the 
%most likely multi-channel scenario is if a second channel exists close to the LyC leaker, near where several apertures that target non-LyC-leaking regions overlap in the source plane.

 The hypotheses we judged to be the strongest both invoke some form of spatially variable \ion{H}{1} with complex morphology, geometry, and/or kinematics which can efficiently diffuse and/or preferentially rescatter a central Ly$\alpha$ peak, likely from the LyC-leaking region. %the existence of a reflective, mirror-like \ion{H}{1} structure in the galaxy's ISM, 
In one scenario (\S\,\ref{sssec:lya_mirror}), dense but perforated \ion{H}{1} in the intermediate space between the LyC-leaking and non-LyC-leaking regions preferentially rescatters a central Ly$\alpha$ peak into our sightline with minimal velocity shift. In the second scenario (\S\,\ref{sssec:disc_lya_diffuse_absorber}), the central Ly$\alpha$ peak actually occupies a much larger area in the source plane than the LyC-leaking region. Although the specific \ion{H}{1} structure and role of absorption are unclear in this explanation, additionally invoking a blueshifted absorber likely offers the simplest explanation and is also consistent with the ubiquitous blueshifted Ly$\alpha$ absorption feature centered around $-350$ km s$^{-1}$. Low-ionization metal absorption lines seen toward the non-LyC-leaking regions also support the existence of such an absorber \citep{mainali+2022}.

% Discuss implications
Another important implication of the work presented here is the direct measurement of the complex, anisotropic relationship between the escape of Ly$\alpha$ and LyC photons, particularly on the small physical scales (hundreds of parsecs and less) probed in the strongly lensed Sunburst Arc. The individual ISM structures that shape Ly$\alpha$ and LyC escape vary at or below this physical scale, which may challenge intuition built from galaxy-integrated properties. The Sunburst Arc is an instructive example of this variability and the small, complex ISM structures we must consider as we develop our understanding of the Ly$\alpha$-LyC connection at sub-galactic scales. The central Ly$\alpha$ peaks without corresponding LyC escape presented in this work are a clear example of this conflict. 

As wide-area surveys such as DECaLS, Pan-STARRS, SDSS, and soon Rubin's Legacy Survey of Space and Time or the Roman Space Telescope's High Latitude Wide Area Survey discover increasing numbers of strongly lensed galaxies, JWST and ELTs will target them with extremely sharp imaging and IFU spectroscopy. We will be able to routinely map distant galaxies down to scales of tens of parsecs or less. Until the discovery of more exceptional objects, the Sunburst Arc offers the highest spatial resolution view of the Ly$\alpha$ and LyC escape processes, and the best opportunity to reconcile disagreements between low and high spatial resolution Ly$\alpha$ and LyC observations. %We suggest that these future observations could challenge the intuitions thus far built from galaxy-integrated properties. %Indeed, we may already observe this in the Sunburst Arc because of the remarkable \textit{positive} correlation measured between $f_{\rm{esc}}^{\rm{LyC}}$ and $v_{\rm{sep}}$.


\section{Author Contributions}

M.R.O. measured the Ly$\alpha$ parameters and $f_{\text{esc}}^{\text{LyC}}$ with advice from K.J.K., M.B.B., and T.E.R.-T. M.R.O. prepared the manuscript with comments and input from all coauthors. M.B.B. obtained the MagE spectra, which were reduced by J.R.R. M.D.G. reduced the HST data. K.S. developed the lens model, created the source plane reconstruction with ray-traced non-LyC-leaking apertures, and calculated the magnification of the MagE apertures. A.N. removed foreground contaminants from the HST imaging to properly estimate $f_{\text{esc}}^{\text{LyC}}$ and added details of this process to the manuscript. This relied upon PSF models created by M.F. with software written by M.D.G. J.G.B. contributed to the background modeling of the HST imaging. M.R.O. created the HST-based Ly$\alpha$ maps based upon software originally written by J.R.R. and SEDs made by G.K. with photometry measured by M.F. G.K. contributed details of the SED modeling to the manuscript.

\section{Access}

The software used to create the results of this paper---and the results themselves---are available freely and publicly on GitHub, including detailed documentation \citep{owens_2024_13888517}.

\begin{acknowledgements}

M.R.O. would like to thank the authors of \citet{chamba+2022} and \citet{knapen+2022} for the helpfulness of their works during the initial development of this manuscript. 
%This work used Jupyter notebooks \citep{jupyter}, and the following Python libraries: Astropy \citep{astropy:2013, astropy:2018, astropy:2022}, Matplotlib \citep{Hunter:2007}, NumPy \citep{harris2020array}, pandas \citep{pandas1-3-1, mckinney-proc-scipy-2010}, Photutils \citep{photutils1-5-0}, Regions \citep{regions0-5}, SciPy \citep{2020SciPy-NMeth}, sigfig, and stsynphot \citep{stsynphot}.
This paper is based on data gathered with the 6.5 meter Magellan Telescopes located at Las Campanas Observatory, Chile. This work is also based, in part, on observations made with the NASA/ESA Hubble Space Telescope obtained from the Space Telescope Science Institute, which is operated by the Association of Universities for Research in Astronomy, Inc., under NASA contract NAS 5–26555. These observations are associated with programs GO-15101 (PI: H. Dahle), GO-15377 (PI: M. Bayliss), GO-15418 (PI: H. Dahle), and GO-15949 (PI: M. Gladders). This work was supported by grants associated with programs GO-15101, GO-15377, GO-15418, and GO-15949 that were awarded by STScI under NASA contract NAS 5-26555. T.E.R.-T. is supported by the Swedish Research Council grant Nr. 2022-04805\_VR. Data presented in this paper were obtained from the Mikulski Archive for Space Telescopes (MAST) at the Space Telescope Science Institute. The specific observations analyzed can be accessed via \dataset[https://doi.org/10.17909/t87g-a816]{https://doi.org/10.17909/t87g-a816}. STScI is operated by the Association of Universities for Research in Astronomy, Inc., under NASA contract NAS5–26555. Support to MAST for these data is provided by the NASA Office of Space Science via grant NAG5–7584 and by other grants and contracts. Support for this work was provided by the National Aeronautics and Space Administration through Chandra Award Number GO1-22081X issued by the Chandra X-ray Center, operated by the Smithsonian Astrophysical Observatory for and on behalf of the National Aeronautics Space Administration under contract NAS8-03060.
This work also makes use of the SAO/NASA Astrophysics Data System's bibliographic services. We prepared this manuscript using the colorblindness simulation tool created by David Nichols\footnote{\href{https://davidmathlogic.com/colorblind}{davidmathlogic.com/colorblind}} to create figures accommodating colorblindness. 

\end{acknowledgements}

\facilities{HST (ACS, WFC3), Magellan:Baade (FIRE, MagE)}

\clearpage

\software{
asciitree,
Astropy \citep{astropy_collaboration+2013, astropy_collaboration+2018, astropy_collaboration+2022}, 
astropy-iers-data \citep{astropy-iers-data},
astropy-healpix,
asttokens,
Beautiful Soup,
Click,
cloudpickle,
Colorama,
Comm,
ContourPy,
cycler,
Dask \citep{dask},
debugpy,
decorator,
executing,
extinction,
Fasteners,
fonttools,
fsspec,
\textsc{Galfit} \citep{peng+2010},
ipykernel,
IPython \citep{ipython},
Jedi,
Jupyter Client,
Jupyter Core,
Jupyter Notebooks \citep{kluyver+2016},
kiwisolver,
lazy\_loader,
llvmlite,
Locket,
Matplotlib \citep{hunter2007,matplotlib_v3.9.2},
matplotlib-inline,
nest-asyncio,
Numba \citep{numba},
Numcodecs,
NumPy \citep{numpy},
packaging,
pandas \citep{pandas_v2.2.2},
Parso,
PartD,
Photutils \citep{photutils_v1.13.0},
Pillow \citep{pillow_v10.4.0},
platformdirs,
prompt\_toolkit,
prospector \citep{johnson+2021},
psutil,
pure\_eval,
PyERFA,
Pygments,
pyparsing,
python-dateutil,
pytz,
pywin32,
PyYAML,
PyZMQ,
Regions \citep{regions_v0.9},
reproject,
scikit-image \citep{scikit-image},
SciPy \citep{scipy},
sigfig,
Six,
Sorted Containers,
Soup Sieve,
stack\_data,
Starburst99 \citep{leitherer+1999},
stsynphot \citep{stsynphot, stsynphot_v1.3.0},
synphot,
toolz,
Tornado,
Traitlets,
tzdata,
wcwidth,
Zarr \citep{zarr_v2.18.2}}

\bibliography{bib}{}
\bibliographystyle{aasjournal}



\end{document}

%% file: f_esc_lyc_measurements_table.tex
\begin{deluxetable}{cccc}[t]

\tablecaption{HST-based properties of the MagE spectra \label{tab:f_esc}}

\tablehead{
	\colhead{Slit} & \colhead{$F_{275}$} & \colhead{$F_{814}$} & \colhead{$f_{\rm{esc}}^{\rm{LyC}}$}
}

\startdata
M5 & $20\pm20$ & $14580\pm70$ & $4\pm5$ \\
M4 & $4\pm20$ & $26760\pm60$ & $0.5\pm3$ \\
M6 & $50\pm20$ & $26700\pm200$ & $6\pm3$ \\
M3 & $40\pm20$ & $26680\pm60$ & $5\pm3$ \\
\hline
M0 & $290\pm30$ & $32580\pm80$ & $46\pm4$ \\
M2 & $370\pm30$ & $39740\pm80$ & $30\pm3$ \\
M7 & $120\pm20$ & $19680\pm60$ & $22\pm4$ \\
M8 & $100\pm30$ & $12900\pm70$ & $20\pm5$ \\
M9 & $150\pm30$ & $21220\pm70$ & $19\pm3$ 
\enddata

\tablecomments{From left to right: slit label, flux in the HST/WFC3 F275W and HST/ACS F814W filters ($10^{-19}$ erg s$^{-1}$ cm$^{-2}$), and $f_{\text{esc}}^{\text{LyC}}$ (\%), all computed according to \S\,\ref{ssec:methods_fesc}.}

\end{deluxetable}

%% file: lya_measurements_table.tex
\begin{deluxetable*}{cllllllll}[ht!]

\tablecaption{Ly$\alpha$ measurements \label{tab:lya_params}}

\tablehead{
	\colhead{Slit} & \colhead{$v_{\rm{sep}}$} & \colhead{FWHM (blue)} & \colhead{FWHM (center)} & \colhead{FWHM (red)} & \colhead{$f_{\rm{min}}/f_{\rm{cont}}$} & \colhead{EW} & \colhead{$f_{\rm{cen}}$} & \colhead{Luminosity}
	\\
	\colhead{} &
	\colhead{[km s$^{-1}$]} &
	\colhead{[km s$^{-1}$]} &
	\colhead{[km s$^{-1}$]} &
	\colhead{[km s$^{-1}$]} &
	\colhead{} &
	\colhead{[\AA]} &
	\colhead{[\%]} &
	\colhead{[$10^{41}$ erg s$^{-1}$]}
}

\startdata
NL & --- & --- & $79_{-9}^{+7}$ & $280_{-9}^{+9}$ & $2.0_{-0.3}^{+0.3}$ & $7.9_{-0.4}^{+0.4}$ & $10.4_{-0.2}^{+0.2}$ & --- \\
L & $331_{-3}^{+3}$ & $202_{-5}^{+4}$ & $79_{-3}^{+3}$ & $261_{-6}^{+9}$ & $16.9_{-0.4}^{+0.3}$ & $25.5_{-0.6}^{+0.6}$ & $26.7_{-0.1}^{+0.1}$ & --- \\
\hline
M5 & --- & --- & $\lesssim 70$ & $256_{-10}^{+9}$ & $0.7_{-0.2}^{+0.3}$ & $4.4_{-0.5}^{+0.5}$ & $7.4_{-0.3}^{+0.3}$ & $2.9_{-0.2}^{+0.2}$ \\
M4 & --- & --- & $\lesssim 80$ & $240_{-8}^{+9}$ & $2.5_{-0.2}^{+0.2}$ & $7.4_{-0.7}^{+0.7}$ & $11.7_{-0.4}^{+0.4}$ & $12.4_{-0.7}^{+0.7}$ \\
M6 & --- & --- & $88_{-9}^{+6}$ & $236_{-5}^{+9}$ & $5.8_{-0.3}^{+0.2}$ & $11.9_{-0.7}^{+0.7}$ & $10.0_{-0.2}^{+0.2}$ & $2.57_{-0.08}^{+0.08}$ \\
M3 & $302_{-7}^{+6}$ & $180_{-10}^{+10}$ & $83_{-7}^{+7}$ & $230_{-6}^{+7}$ & $11.5_{-0.5}^{+0.4}$ & $24_{-1}^{+1}$ & $24.5_{-0.2}^{+0.2}$ & $14.5_{-0.3}^{+0.3}$ \\
\hline
M0 & $355_{-4}^{+4}$ & $175_{-5}^{+6}$ & $89_{-3}^{+3}$ & $244_{-4}^{+4}$ & $15.7_{-0.3}^{+0.3}$ & $23.5_{-0.7}^{+0.8}$ & $19.8_{-0.1}^{+0.1}$ & $1120_{-10}^{+10}$\tablenotemark{a}\\
M2 & $342_{-10}^{+8}$ & $180_{-20}^{+10}$ & $70_{-5}^{+7}$ & $250_{-20}^{+20}$ & $18.3_{-0.8}^{+0.7}$ & $23_{-1}^{+1}$ & $25.6_{-0.2}^{+0.2}$ & $24.9_{-0.4}^{+0.4}$ \\
M7 & $313_{-4}^{+4}$ & $202_{-5}^{+6}$ & $76_{-3}^{+3}$ & $244_{-4}^{+4}$ & $17.1_{-0.3}^{+0.3}$ & $25.6_{-0.7}^{+0.7}$ & $29.8_{-0.2}^{+0.2}$ & $25.4_{-0.2}^{+0.2}$ \\
M8 & $326_{-4}^{+4}$ & $196_{-7}^{+7}$ & $80_{-5}^{+4}$ & $258_{-5}^{+9}$ & $18.6_{-0.5}^{+0.5}$ & $29_{-1}^{+1}$ & $21.4_{-0.1}^{+0.1}$ & $31.5_{-0.3}^{+0.3}$ \\
M9 & $328_{-7}^{+7}$ & $200_{-10}^{+10}$ & $85_{-5}^{+4}$ & $255_{-6}^{+6}$ & $14.5_{-0.5}^{+0.5}$ & $25_{-1}^{+2}$ & $26.4_{-0.3}^{+0.2}$ & $18.1_{-0.3}^{+0.3}$ 
\enddata

\tablecomments{From left to right: slit label, peak separation between the redshifted and blueshifted Ly$\alpha$ peaks (km s$^{-1}$), FWHM of the blueshifted, central, and redshifted Ly$\alpha$ peaks (km s$^{-1}$), respectively, ratio between the `minimum' flux density between the redshifted and blueshifted Ly$\alpha$ peaks and the local continuum flux density, rest-frame Ly$\alpha$ equivalent width ({\AA}), central escape fraction (\%), and magnification-corrected Ly$\alpha$ luminosity (10$^{41}$ erg s$^{-1}$). Because the deconvolved FWHMs of the central Ly$\alpha$ peaks of slits M4 and M5 were not significantly greater than the instrumental line spread function FWHM ($\sim55$ km s$^{-1}$), we quote the 84th percentiles of those measurements as an upper bound on the intrinsic FWHM of their central Ly$\alpha$ peaks.}

\tablenotetext{a}{Slit M0's observation was taken through thin cloud cover that prevented an accurate fluxing, so its significantly larger luminosity is not an accurate estimate. We do not include this data point in any figures or when estimating any correlations involving the Ly$\alpha$ luminosity. See Table \ref{tab:mage_log} for more information about the observation.}

\end{deluxetable*}

%% file: lya_measurements_statistical_correlations_table.tex
\begin{deluxetable}{lrr}

\tablecaption{Correlations between Ly$\alpha$ and LyC parameters \label{tab:param_correlations}}

\tablehead{
	\colhead{} & \colhead{$r$} & \colhead{$\tau$}
}

\startdata
FWHM (b) - $v_{\rm{sep}}$ & $-0.4_{-0.2}^{+0.2}$ & $-0.2_{-0.2}^{+0.2}$ \\
FWHM (c) - $v_{\rm{sep}}$ & $0.1_{-0.3}^{+0.4}$ & $0.0_{-0.2}^{+0.3}$ \\
FWHM (r) - $v_{\rm{sep}}$ & $0.4_{-0.2}^{+0.2}$ & $0.3_{-0.2}^{+0.2}$ \\
$f_{\rm{min}}/f_{\rm{cont}}$ - $v_{\rm{sep}}$ & $0.4_{-0.2}^{+0.1}$ & $0.1_{-0.1}^{+0.2}$ \\
Ly$\alpha$ EW - $v_{\rm{sep}}$ & $-0.2_{-0.2}^{+0.2}$ & $-0.1_{-0.2}^{+0.2}$ \\
$f_{\rm{cen}}$ - $v_{\rm{sep}}$ & $-0.5_{-0.1}^{+0.1}$ & $-0.3_{-0.1}^{+0.1}$ \\
Ly$\alpha$ $L$ - $v_{\rm{sep}}$ & $0.4_{-0.2}^{+0.2}$ & $0.2_{-0}^{+0.2}$ \\
$f_{\rm{esc}}^{\rm{LyC}}$ - $v_{\rm{sep}}$ & $0.93_{-0.05}^{+0.03}$ & $0.7_{-0.1}^{+0.1}$ \\
FWHM (c) - FWHM (b) & $-0.3_{-0.3}^{+0.3}$ & $-0.2_{-0.2}^{+0.3}$ \\
FWHM (r) - FWHM (b) & $0.5_{-0.4}^{+0.2}$ & $0.3_{-0.3}^{+0.2}$ \\
$f_{\rm{min}}/f_{\rm{cont}}$ - FWHM (b) & $0.1_{-0.3}^{+0.3}$ & $0.0_{-0.2}^{+0.2}$ \\
Ly$\alpha$ EW - FWHM (b) & $0.5_{-0.2}^{+0.2}$ & $0.4_{-0.2}^{+0.2}$ \\
$f_{\rm{cen}}$ - FWHM (b) & $0.5_{-0.2}^{+0.2}$ & $0.4_{-0.2}^{+0.2}$ \\
Ly$\alpha$ $L$ - FWHM (b) & $0.2_{-0.3}^{+0.3}$ & $0.2_{-0.2}^{+0.4}$ \\
$f_{\rm{esc}}^{\rm{LyC}}$ - FWHM (b) & $-0.71_{-0.06}^{+0.08}$ & $-0.5_{-0.1}^{+0.1}$ \\
FWHM (r) - FWHM (c) & $-0.2_{-0.2}^{+0.2}$ & $-0.2_{-0.2}^{+0.2}$ \\
$f_{\rm{min}}/f_{\rm{cont}}$ - FWHM (c) & $0.4_{-0.2}^{+0.2}$ & $0.1_{-0.2}^{+0.1}$ \\
Ly$\alpha$ EW - FWHM (c) & $0.5_{-0.2}^{+0.2}$ & $0.2_{-0.2}^{+0.2}$ \\
$f_{\rm{cen}}$ - FWHM (c) & $0.3_{-0.2}^{+0.2}$ & $0.1_{-0.2}^{+0.1}$ \\
Ly$\alpha$ $L$ - FWHM (c) & $0.2_{-0.2}^{+0.2}$ & $0.1_{-0.2}^{+0.1}$ \\
$f_{\rm{esc}}^{\rm{LyC}}$ - FWHM (c) & $0.51_{-0.08}^{+0.08}$ & $0.28_{-0.1}^{+0.06}$ \\
$f_{\rm{min}}/f_{\rm{cont}}$ - FWHM (r) & $-0.1_{-0.2}^{+0.2}$ & $0.1_{-0.1}^{+0.1}$ \\
Ly$\alpha$ EW - FWHM (r) & $-0.1_{-0.2}^{+0.2}$ & $0.1_{-0.1}^{+0.1}$ \\
$f_{\rm{cen}}$ - FWHM (r) & $-0.1_{-0.2}^{+0.2}$ & $0.0_{-0.1}^{+0.1}$ \\
Ly$\alpha$ $L$ - FWHM (r) & $0.3_{-0.3}^{+0.2}$ & $0.3_{-0.2}^{+0.1}$ \\
$f_{\rm{esc}}^{\rm{LyC}}$ - FWHM (r) & $0.18_{-0.09}^{+0.09}$ & $0.28_{-0.1}^{+0.06}$ \\
Ly$\alpha$ EW - $f_{\rm{min}}/f_{\rm{cont}}$ & $0.965_{-0.01}^{+0.009}$ & $0.71_{-0.07}^{+0.07}$ \\
$f_{\rm{cen}}$ - $f_{\rm{min}}/f_{\rm{cont}}$ & $0.898_{-0.009}^{+0.009}$ & $0.56_{-0.04}^{+0.04}$ \\
Ly$\alpha$ $L$ - $f_{\rm{min}}/f_{\rm{cont}}$ & $0.90_{-0.01}^{+0.01}$ & $0.79_{-0.07}^{+0.07}$ \\
$f_{\rm{esc}}^{\rm{LyC}}$ - $f_{\rm{min}}/f_{\rm{cont}}$ & $0.74_{-0.07}^{+0.06}$ & $0.61_{-0.06}^{+0.1}$ \\
$f_{\rm{cen}}$ - Ly$\alpha$ EW & $0.91_{-0.01}^{+0.01}$ & $0.60_{-0.07}^{+0.04}$ \\
Ly$\alpha$ $L$ - Ly$\alpha$ EW & $0.85_{-0.02}^{+0.02}$ & $0.71_{-0.07}^{+0.07}$ \\
$f_{\rm{esc}}^{\rm{LyC}}$ - Ly$\alpha$ EW & $0.60_{-0.08}^{+0.07}$ & $0.3_{-0.1}^{+0.1}$ \\
Ly$\alpha$ $L$ - $f_{\rm{cen}}$ & $0.79_{-0.01}^{+0.01}$ & $0.6_{-0}^{+0}$ \\
$f_{\rm{esc}}^{\rm{LyC}}$ - $f_{\rm{cen}}$ & $0.50_{-0.08}^{+0.07}$ & $0.33_{-0.1}^{+0.06}$ \\
$f_{\rm{esc}}^{\rm{LyC}}$ - Ly$\alpha$ $L$ & $0.50_{-0.07}^{+0.07}$ & $0.33_{-0.1}^{+0.06}$ 
\enddata

\tablecomments{Statistical correlations between the Ly$\alpha$ parameters and $f_{\rm{esc}}^{\rm{LyC}}$. From left to right: the parameter pair, Pearson correlation coefficient $r$, and type `b' Kendall rank correlation coefficient $\tau$. The minimal number of data points (no more than 11 for any pair of parameters) means there are not many unique values of $\tau$, which causes many of the listed values and uncertainties to be similar, or in extreme cases for high SNR parameter measurements, for the 16th and 84th percentiles listed to be the same value as the median.}

\end{deluxetable}

%% file: seeing_simulation_measurements_table.tex
\begin{deluxetable*}{r|lll|lll}
\tablecaption{Simulated aperture flux changes due to atmospheric seeing \label{tab:seeing_simulation}}

\tablehead{
	\colhead{Slit} & 
	\colhead{} &
	\colhead{F390W} &
	\colhead{} &
	\colhead{} &
	\colhead{F555W} & 
	\colhead{}
	\\
	\colhead{} &
	\colhead{Unconvolved} &
	\colhead{Convolved} &
	\colhead{Ratio} & 
	\colhead{Unconvolved} &
	\colhead{Convolved} &
	\colhead{Ratio}
	\\
	\colhead{} &
	\colhead{[$e^-$ s$^{-1}$]} &
	\colhead{[$e^-$ s$^{-1}$]} &
	\colhead{} &
	\colhead{[$e^-$ s$^{-1}$]} &
	\colhead{[$e^-$ s$^{-1}$]} &
	\colhead{}
}

\startdata
M5 & $0.82\pm0.03$ & $0.86\pm0.03$ & $1.05\pm0.05$ & $-0.15\pm0.04$ & $-0.21\pm0.04$ & $1.4\pm0.5$ \\
M4 & $1.34\pm0.03$ & $1.42\pm0.03$ & $1.06\pm0.03$ & $0.04\pm0.04$ & $0.35\pm0.04$ & $10\pm10$ \\
M6 & $1.66\pm0.03$ & $1.80\pm0.03$ & $1.08\pm0.02$ & $0.49\pm0.04$ & $0.81\pm0.04$ & $1.6\pm0.2$ \\
M3 & $1.51\pm0.04$ & $1.61\pm0.04$ & $1.06\pm0.04$ & $-0.24\pm0.08$ & $0.25\pm0.08$ & $-1.0\pm0.5$ \\
\hline
M0 & $5.90\pm0.06$ & $4.40\pm0.06$ & $0.75\pm0.01$ & $3.6\pm0.1$ & $2.8\pm0.1$ & $0.78\pm0.04$ \\
M2 & $6.63\pm0.07$ & $5.10\pm0.07$ & $0.77\pm0.01$ & $5.2\pm0.1$ & $4.0\pm0.1$ & $0.77\pm0.03$ \\
M7 & $3.22\pm0.04$ & $2.50\pm0.04$ & $0.78\pm0.01$ & $2.18\pm0.07$ & $1.70\pm0.07$ & $0.78\pm0.04$ \\
M8 & $2.49\pm0.02$ & $2.21\pm0.02$ & $0.89\pm0.01$ & $1.98\pm0.04$ & $1.79\pm0.04$ & $0.90\pm0.03$ \\
M9 & $3.02\pm0.03$ & $2.43\pm0.03$ & $0.80\pm0.01$ & $2.27\pm0.06$ & $1.46\pm0.06$ & $0.64\pm0.03$
\enddata

\tablecomments{From left to right: slit, flux inside the slit of the unconvolved and simulated seeing-convolved F390W-based Ly$\alpha$ map, the ratio between the two fluxes, and likewise for the F555W-based Ly$\alpha$ map. Ratios $>$ 1 indicate the flux in the aperture increased after the convolution, and ratios $<$ 1 indicate the flux in the aperture decreased after the convolution. The convolution used a 2-dimensional Gaussian kernel of the combined effect of the time-weighted seeing conditions and airmasses (\S\,\ref{sssec:disc_lya_seeing}).}
\end{deluxetable*}

%% file: lya_best_fit_model_parameters_table.tex
\begin{deluxetable}{lcccc}

\rotate

\tablecaption{Ly$\alpha$ modeling best-fit parameters \label{tab:fit_results}}

\tablehead{
	\colhead{Slit} & \colhead{Blue peak} & \colhead{Red peak} & \colhead{Central peak} & \colhead{Continuum}
}

\startdata
NL & $-$ / $-$ / $-$ / $-$ & $3.74_{-0.07}^{+0.07}$ / $127_{-5}^{+5}$ / $210_{-4}^{+4}$ / $3.5_{-0.3}^{+0.4}$ & $2.3_{-0.3}^{+0.3}$ / $60_{-5}^{+5}$ / $41_{-3}^{+2}$ / $-$ & $1.17_{-0.02}^{+0.02}$ \\
L & $6.58_{-0.07}^{+0.07}$ / $-99_{-2}^{+2}$ / $89_{-2}^{+2}$ / $-0.0002_{-0.003}^{+0.0002}$ & $5.7_{-0.1}^{+0.1}$ / $157_{-6}^{+5}$ / $210_{-2}^{+3}$ / $5.3_{-0.8}^{+0.9}$ & $22.6_{-0.4}^{+0.3}$ / $79_{-1}^{+1}$ / $41_{-1}^{+1}$ / $-$ & $1.34_{-0.02}^{+0.02}$ \\
M5 & $-$ / $-$ / $-$ / $-$ & $1.91_{-0.06}^{+0.06}$ / $121_{-4}^{+5}$ / $204_{-7}^{+7}$ / $5.1_{-0.7}^{+1}$ & $0.6_{-0.2}^{+0.3}$ / $40_{-10}^{+20}$ / $30_{-10}^{+10}$ / $-$ & $0.92_{-0.02}^{+0.02}$ \\
M4 & $-$ / $-$ / $-$ / $-$ & $1.93_{-0.07}^{+0.07}$ / $128_{-4}^{+4}$ / $194_{-6}^{+6}$ / $6_{-1}^{+2}$ & $1.7_{-0.1}^{+0.1}$ / $44_{-5}^{+6}$ / $36_{-4}^{+5}$ / $-$ & $0.69_{-0.02}^{+0.02}$ \\
M6 & $-$ / $-$ / $-$ / $-$ & $3.15_{-0.07}^{+0.1}$ / $175_{-5}^{+3}$ / $193_{-4}^{+4}$ / $6_{-2}^{+1}$ & $5.1_{-0.2}^{+0.2}$ / $95_{-6}^{+3}$ / $45_{-3}^{+2}$ / $-$ & $0.87_{-0.02}^{+0.02}$ \\
M3 & $3.56_{-0.09}^{+0.08}$ / $-66_{-6}^{+6}$ / $81_{-4}^{+4}$ / $-0.0004_{-0.004}^{+0.0004}$ & $2.96_{-0.07}^{+0.07}$ / $168_{-4}^{+3}$ / $184_{-3}^{+4}$ / $5.0_{-0.8}^{+0.9}$ & $7.5_{-0.3}^{+0.2}$ / $77_{-2}^{+2}$ / $42_{-2}^{+2}$ / $-$ & $0.65_{-0.01}^{+0.01}$ \\
\hline
M0 & $55.4_{-0.9}^{+0.9}$ / $-105_{-2}^{+2}$ / $79_{-2}^{+2}$ / $-0.0009_{-0.002}^{+0.0008}$ & $48.5_{-0.7}^{+0.8}$ / $186_{-3}^{+2}$ / $202_{-3}^{+3}$ / $6.5_{-0.7}^{+0.7}$ & $208_{-2}^{+2}$ / $94.5_{-0.8}^{+0.7}$ / $46.6_{-1}^{+0.9}$ / $-$ & $13.3_{-0.2}^{+0.2}$ \\
M2 & $3.6_{-1}^{+0.2}$ / $-102_{-7}^{+70}$ / $85_{-5}^{+40}$ / $-0.05_{-3}^{+0.05}$ & $3.8_{-0.2}^{+0.2}$ / $160_{-10}^{+10}$ / $197_{-5}^{+6}$ / $4_{-1}^{+2}$ & $18.1_{-0.8}^{+0.5}$ / $84_{-1}^{+2}$ / $38_{-1}^{+3}$ / $-$ & $0.99_{-0.03}^{+0.03}$ \\
M7 & $5.22_{-0.07}^{+0.07}$ / $-103_{-3}^{+3}$ / $89_{-2}^{+2}$ / $-0.0008_{-0.003}^{+0.0008}$ & $4.01_{-0.05}^{+0.06}$ / $150_{-3}^{+2}$ / $203_{-3}^{+2}$ / $7.4_{-0.9}^{+0.8}$ & $16.1_{-0.2}^{+0.2}$ / $67.7_{-0.8}^{+0.7}$ / $40.6_{-1}^{+0.9}$ / $-$ & $0.94_{-0.01}^{+0.01}$ \\
M8 & $4.77_{-0.07}^{+0.07}$ / $-91_{-3}^{+3}$ / $87_{-3}^{+3}$ / $-0.0006_{-0.003}^{+0.0006}$ & $4.08_{-0.07}^{+0.1}$ / $168_{-7}^{+3}$ / $212_{-3}^{+4}$ / $7_{-2}^{+1}$ & $16.6_{-0.3}^{+0.2}$ / $86_{-2}^{+1}$ / $42_{-2}^{+1}$ / $-$ & $0.89_{-0.02}^{+0.02}$ \\
M9 & $2.89_{-0.08}^{+0.08}$ / $-100_{-4}^{+5}$ / $88_{-4}^{+5}$ / $-0.001_{-0.004}^{+0.001}$ & $2.42_{-0.05}^{+0.06}$ / $170_{-3}^{+2}$ / $213_{-5}^{+5}$ / $8_{-2}^{+2}$ & $9.6_{-0.2}^{+0.2}$ / $80_{-1}^{+1}$ / $43_{-1}^{+1}$ / $-$ & $0.66_{-0.02}^{+0.02}$ 
\enddata

\tablecomments{Best-fit Ly$\alpha$ parameters of the magnification-uncorrected, rest-frame Ly$\alpha$ profiles. From left to right: slit label, best-fit parameters for the blueshifted peak, redshifted peak, and central peak, presented as $\alpha$/$\mu$/$\sigma$/$\omega$, as described in \S\,\ref{sssec:methods_lya_fwhm_vsep}. Apart from the stacked spectra, $\alpha$ has units of $10^{-16}$ erg s$^{-1}$ cm$^{-2}$ {\AA}$^{-1}$, and $\mu$ and $\sigma$ have units of km s$^{-1}$.}

\tablenotetext{a}{The fitted amplitudes of slit M0 are significantly different from the other observations because this observation was taken through cloud cover that prevented an accurate fluxing. See Table \ref{tab:mage_log} for more information about the observing conditions.}

\end{deluxetable}